%% file: main.tex
\documentclass[runningheads]{llncs}

\usepackage{cite}
\usepackage[T1]{fontenc}
\usepackage{graphicx}
\usepackage{subcaption}
\usepackage{amsmath}
\usepackage[ruled]{algorithm}
\usepackage{algpseudocode}
\usepackage{amssymb}
\usepackage{multirow}
\usepackage{xcolor}
\usepackage[normalem]{ulem}
\usepackage{url}

\renewcommand{\sout}[1]{}
\renewcommand{\textcolor}[2]{#2}

\begin{document}
\title{Modeling and Scheduling of Fusion Patterns in Autonomous Driving Systems}

\subtitle{\small (Extended Version) \vspace{-6mm}}
\titlerunning{Modeling and \textcolor{blue}{Scheduling} of Fusion Patterns in ADS}

\author{Hoora Sobhani\orcidID{0009-0009-6611-9374} \and
Hyoseung Kim\orcidID{0000-0002-8553-732X}}
\authorrunning{H. Sobhani et al.}
\institute{University of California, Riverside (UCR)\\
\email{\{hsobh002, hyoseung\}@ucr.edu}}
\maketitle              
\begin{abstract}
In Autonomous Driving Systems (ADS), Directed Acyclic Graphs (DAGs) are widely used to model complex data dependencies and inter-task communication. However, existing DAG scheduling approaches oversimplify data fusion tasks by assuming fixed triggering mechanisms, failing to capture the diverse fusion patterns found in real-world ADS software stacks. In this paper, we propose a systematic framework for analyzing various fusion patterns and their performance implications in ADS. Our framework models three distinct fusion task types: timer-triggered, wait-for-all, and immediate fusion, which comprehensively represent real-world fusion behaviors. Our Integer Linear Programming (ILP)-based approach enables \sout{a simultaneous}\textcolor{blue}{an} optimization of multiple real-time performance metrics, including reaction time, time disparity, age of information, and response time, while generating deterministic offline schedules directly applicable to real platforms. Evaluation using real-world ADS case studies, Raspberry Pi implementation, and randomly generated DAGs demonstrates that our framework handles diverse fusion patterns beyond the scope of existing work, and achieves substantial performance improvements in comparable scenarios.
\end{abstract}
\section{Introduction}\label{sec:intro}

Research in Autonomous Driving Systems (ADS) has attracted strong interest from academia and industry, aiming to develop safe and reliable autonomous vehicles. \sout{ADS combines technologies like computer vision, machine learning, and sensor fusion into a complex, real-time embedded system.} The ADS software stack consists of tightly interconnected components with intricate data dependencies, interacting through a deep processing pipeline \cite{rtss2021challenge, kim2013parallel, zhu2020know}. These components include sensors, algorithms, and actuators, all working together under strict timing constraints to ensure safety. 
They are often represented as nodes in Directed Acyclic Graphs (DAGs), capturing the system’s workflow, task dependencies, and inter-task communication.

DAGs have been instrumental in studying task scheduling and resource allocation in ADS, especially to improve key performance metrics like end-to-end latency, sensor time disparity, data freshness, and responsiveness in safety-critical, real-time environments. However, existing DAG scheduling models often fall short in fully capturing the diverse behaviors of real-world ADS, such as Autoware~\cite{autoware}. One major challenge is accurately modeling data \textit{fusion nodes}, which aggregate multiple inputs to generate outputs. In ADS, numerous sensors operate at different sampling rates, passing their data through these fusion nodes. \textcolor{blue}{Dealing with the varying sampling rates of sensors is inevitable due to the limited flexibility of sensor hardware (e.g., lidar and camera cannot choose arbitrary sampling periods). Enforcing rate uniformity may cause all tasks to run with the shortest period or with the GCD of all periods. In addition,} \sout{Beyond the multi-rate data arrivals of sensors,} the fusion nodes add complexity as they can be triggered either by their own timers or by external events with unpredictable arrival times. This complexity deepens in event-triggered cases, where any of the inputs to a fusion node can independently initiate its activation. Such variability in activation timing and patterns complicates modeling and scheduling, as traditional DAG models often assume more uniform or simpler task activation mechanisms. For instance, authors in \cite{toba2024deadline, yano2023deadline, teper2022end} assumed that a fusion node is triggered by a single dominant input edge, while authors in \cite{sun2023real} assumed all nodes, including fusion nodes, are triggered solely by timers. 
\textcolor{blue}{While these models may fit certain applications, existing work focuses on only one type of fusion behavior, overlooking the various types found in real-world ADS such as Autoware.}
\sout{While these fit certain real-world Autoware components like \texttt{tracking\_object\_merger} and \texttt{ndt\_scan\_matcher}, they fail to capture more dynamic behaviors like that of \texttt{traffic\_light\_occlusion\_predictor}, where any incoming edge can trigger the fusion node’s execution. Moreover, existing work focuses on only one type of fusion behavior, overlooking that various types may coexist in an ADS application.}

\sout{Beyond limitations in modeling fusion nodes, existing studies face other critical gaps. A major limitation is their tendency to optimize a specific real-time metric for individual chains,}\textcolor{blue}{Beyond fusion node limitations, existing studies typically optimize individual chains,} rather than considering the entire DAG holistically with multiple metrics~\cite{teper2022end, xu2022aoi, sun2023real}. \sout{Most approaches target chain response time, time disparity, or reaction time in isolation, or analyze paths from one sensor to an actuator while overlooking multiple sensors collectively contributing to an actuator’s output. Such limitations can lead to suboptimal system performance when chain interactions are not accounted for. Additionally, many studies, even those focusing on individual chains, support only certain types of cause-effect chains. For instance, several studies focus solely on \textit{multi-rate cause-effect chains}, where all tasks are triggered by timers with no event-triggered tasks~\cite{sun2023real, becker2016synthesizing}. More flexible models}\textcolor{blue}{Even studies that} include both timer- and event-triggered tasks ~\cite{teper2022end, teper2024end, gunzel2023equivalence}\sout{, but even these} fail to capture real-world ADS configurations like \sout{branch-and-fusion path to actuators}\textcolor{blue}{branch-then-fusion paths (where multiple outputs fork from a common node and later converge into a single node)}
and diverse triggering mechanisms\sout{that drive data propagation through the DAG}. Addressing these gaps is essential to reflect the dynamic and interconnected nature of ADS workflows.


In this paper, we offer a systematic framework for analyzing various fusion patterns in ADS and their impact on real-time performance. To explore the best achievable performance under different fusion strategies, we formulate the DAG scheduling problem and diverse fusion patterns as Integer Linear Programming (ILP) constraints, 
as ILP allows us to obtain optimal schedules\sout{ for deterministic execution} while offering extensibility \sout{and expressiveness} in constraint modeling.
\sout{With this approach, we navigate the optimal solution space, enabling}\textcolor{blue}{Our approach enables} a quantitative comparison of how different fusion strategies affect key performance metrics such as worst-case response time (WCRT), maximum reaction time (MRT), maximum time disparity (MTD), and peak age of information (PAoI), while also uncovering trade-offs of different fusion nodes in ADS design. 
The optimal schedule generated \sout{offline by our ILP-based approach includes both resource allocation (task-to-core mapping) and the timing scheme for all task instances within the DAG; hence, this}\textcolor{blue}{by our framework} can be directly applicable to real-world ADS with static non-preemptive scheduling, such as NVIDIA's STM~\cite{davies2025stm}. We evaluate our framework through comparison against state-of-the-art methods, implementation on a Raspberry Pi, and experiments with randomly generated complex DAGs. Experimental results show that our framework successfully handles various fusion patterns beyond the scope of prior work and achieves substantial improvements over existing methods in comparable scenarios.




\section{Related Work}
\label{sec:related}
\sout{This section reviews the literature on DAG scheduling and data fusion modeling in real-time systems, focusing on ADS applications. Prior studies address various environments such as real-time operating systems \cite{xu2022aoi}, middleware like ROS 2 \cite{sun2023real}, and architectures like AUTOSAR \cite{feiertag2009compositional, gunzel2023compositional}, applied to both single-core \cite{teper2022end} and multi-core platforms \cite{teper2024end}. Despite differences in environment and platform setup, they all structure tasks and dependencies as chains or DAGs, tackling complex interdependencies to improve performance metrics in systems with stringent timing requirements.

A key factor in categorizing these studies is}\textcolor{blue}{Prior work on DAG scheduling in real-time systems, particularly for ADS applications, can be categorized by} the type of chains used to construct DAG models, which differ in data communication and task-triggering \textcolor{blue}{mechanisms}. Generally, a chain is a sequence of tasks\sout{ communicating in an order}, where each task depends on data from its predecessor. Tasks within a chain can be triggered by either \textit{timers} or \textit{events}. A widely adopted\sout{ chain} model\sout{ in ADS applications} is the \textbf{cause-effect chain}, traditionally representing\sout{ a sequence of} multi-rate real-time tasks\sout{, each} triggered solely by timers\sout{ set} at different rates \cite{becker2016synthesizing, verucchi2020latency, tang2022comparing}. These chains have evolved to include event-triggered tasks, enabling responses to specific events rather than relying entirely on timers \cite{tang2023reaction, tang2023optimizing, teper2024end, toba2024deadline, yano2023deadline}. We refer to the former as \textbf{multi-rate cause-effect chains} and the latter as \textbf{enhanced cause-effect chains} throughout this section.

Many studies on cause-effect chains aim to minimize metrics like maximum reaction time (MRT) and maximum data age (MDA)\textcolor{blue}{, which are typically used to evaluate system responsiveness and data freshness, respectively}. For example, Becker et al. eliminated data paths in a DAG that exceed age constraints to improve MRT and MDA \cite{becker2016synthesizing, becker2017generic}. A common goal across these studies is to model task communication across the DAG and explore solutions that streamline data flow and fusion for faster responsiveness to sensor data and improved data freshness. For multi-rate cause-effect chains, Tang et al. \cite{tang2022comparing} compared communication paradigms (implicit, LET, DBP), while Maia et al. \cite{maia2024reducing} explored strategies to adjust communication intervals.  Saidi et al. \cite{saidi2017automatic} introduced a graph transformation approach to establish execution order, and Verucchi et al. \cite{verucchi2020latency} presented a method for converting multi-rate DAG chains into single-rate DAGs. Beyond the focus on upper-bounding MRT and MDA \cite{durr2019end, li2024priority}, Jiang et al. \cite{jiang2023analysis} examined maximum time disparity (MTD)\textcolor{blue}{—typically used to measure deviations in sensor sampling times—}between multi-sensor data, emphasizing the need for accurate synchronization for fusion. For enhanced cause-effect chains, Tang et al. \cite{tang2023reaction} reduced MRT via buffer limits and data refreshing, and later proposed a dynamic priority inheritance and buffer manipulation protocol to lower MRT and MDA in their sporadic variant of these chains \cite{tang2023optimizing}. \textcolor{blue}{Regarding upper-bounding MTD alongside end-to-end latency, Sun et al. \cite{sun2025jointly} proposed a mechanism to select which sensor data is used by an actuator and fused along the path from sensor to actuator, assuming fusion nodes are triggered only when a new data arrives.}

Data communication and fusion in cause-effect chains are also addressed in ROS 2-related studies. Li et. al.~\cite{li2023worst} explored the potential and limitations of ROS 2 message synchronizer in multi-sensor data fusion, while Sun et. al.~\cite{sun2023seam} proposed a novel message synchronization policy for ROS 2 to improve MTD when fusing multi-sensor data. Saito et. al.~\cite{saito2016priority} proposed a priority-based message transmission and a synchronization node to address MTD of sensor data. Later, in \cite{saito2018rosch}, they introduced a synchronization system that buffers the highest-rate sensor to align periods, converting multi-rate DAGs into single-rate ones. Sun et al.~\cite{sun2023real} proposed an ILP-based model to optimize MRT and MDA by reducing redundant workload and unnecessary messages. For enhanced cause-effect chains, the model in \cite{gunzel2023compositional,  gunzel2021timing,  gunzel2023equivalence, teper2024end, teper2022end} stands as one of the most advanced implementations. They analyzed MRT and MDA within a single-threaded executor setup by classifying intra- and inter-task data communication patterns \cite{teper2022end} and later demonstrated cases in which MRT and MDA can be equivalent \cite{gunzel2023equivalence}. This analysis was extended to multiple single-threaded executors, incorporating additional communication paradigms \cite{teper2024end}. Yano et al. \cite{yano2024work} proposed a scheduling model that decomposes the Autoware DAG into multi-deadline sub-DAGs, aiming to bypass challenges related to synchronization, queue consumption patterns, and intricate data dependencies. 


\sout{Investigations across these studies into various task-triggering options and data fusion patterns reveal that existing models, even the most advanced ones, fail to fully capture certain real-world ADS tasks. For instance, ADS software with branch-and-fusion tasks or fusion nodes triggered by the arrival of any input cannot be accurately represented by these approaches}\textcolor{blue}{In summary, existing approaches do not fully capture the complexity of real-world ADS that feature multiple triggering options for fusion nodes and branch-then-fusion paths. While each prior work addresses specific fusion patterns, none provides a comprehensive framework that handles all fusion behaviors found in practice (more details in Sec.~\ref{sec:model_fusion}). Moreover, integrating these approaches to analyze complete ADS is non-trivial, since each work employs different assumptions, abstractions, and analytical approaches, making them often incompatible with one another.}
This highlights a gap in modeling fusion nodes within complex ADS structures, \sout{particularly when tasks exhibit diverse triggering behaviors. Therefore, in this paper, we aim to propose a systematic and adaptable framework featuring a flexible DAG model designed to analyze various fusion patterns and task-triggering mechanisms in ADS. }\textcolor{blue}{which motivates our work.}

\section{Background and System Model}\label{sec:bg&sys}
\sout{In this section, we briefly overview the ADS software stack and then present our system model.}

\subsection{Overview of ADS software stack}\label{sec:dag_ads}
\begin{figure}[t]
\centerline{\includegraphics[width=0.65\linewidth]{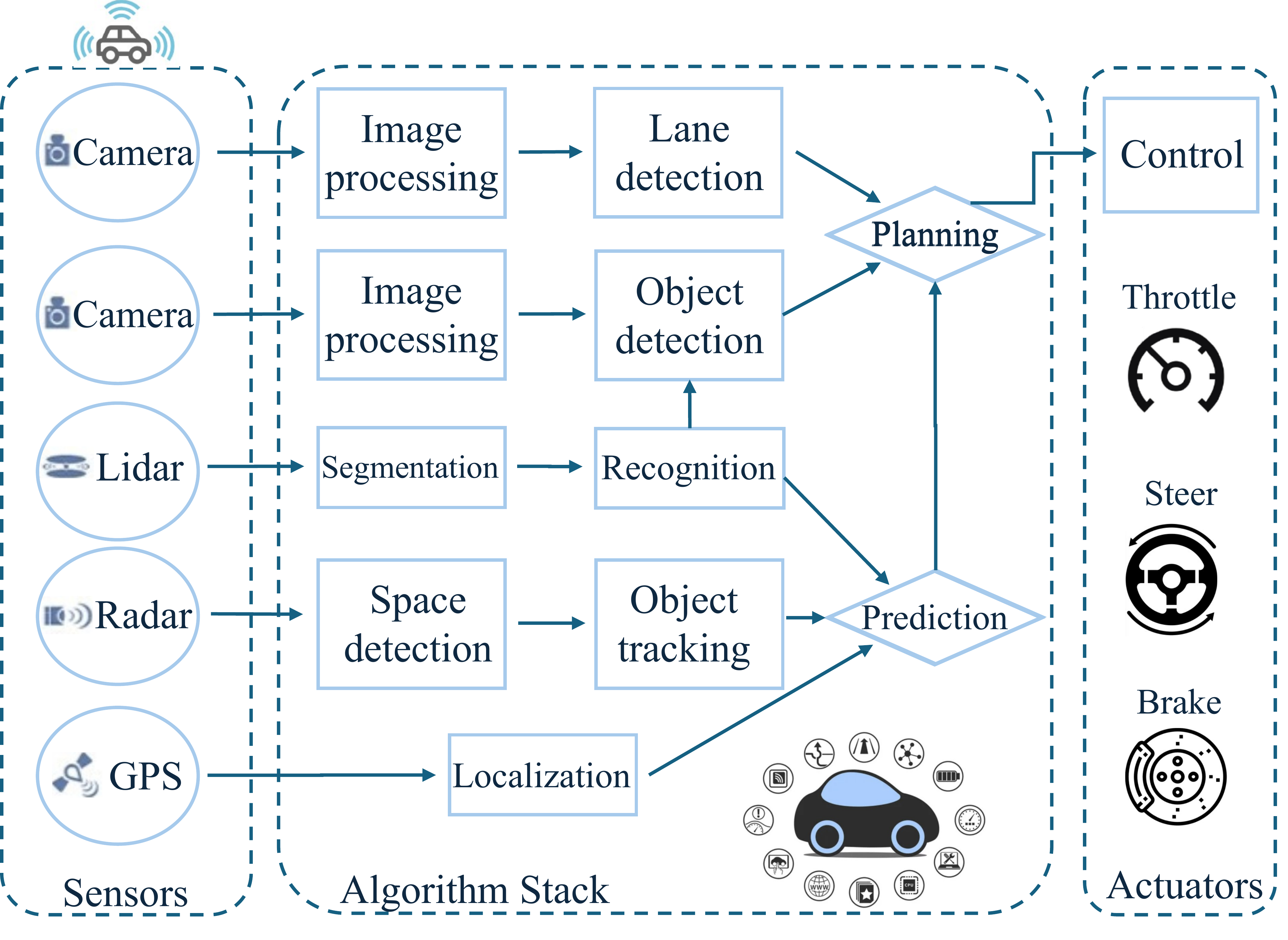}}\vspace{-3mm}
\caption{Holistic overview of the ADS software stack, inspired by \cite{xu2022aoi, kuhse2024sync}.}
\label{fig:ads}
\end{figure}

ADS consists of tightly integrated software components, each essential to the vehicle’s safe operation. As shown in Fig.~\ref{fig:ads}, these components can be categorized into three main classes: sensors, algorithm stack, and control systems (actuators). \textit{Sensors}, including LiDAR, radar, cameras, GPS, and others, gather real-time data from the environment, which serves as input for the system's perception. The \textit{algorithm stack}, which includes object and lane detection, localization, prediction, planning, etc., processes sensor data and interprets it to real-time situational-aware decisions. Finally, \textit{actuators} execute these decisions to control vehicle dynamics, including steering, acceleration, and braking. 


Communication between these interdependent software components happens through a deep processing pipeline. Data sampled by sensors is processed and, at various points along the pipeline, combined with data from other sensors or outputs from intermediate components in the algorithm stack, such as prediction or planning nodes, as shown in Fig.~\ref{fig:ads}. These nodes, called fusion nodes, integrate multiple inputs to produce outputs, with their activation timing and triggering patterns varying based on the ADS application's specifications. The interaction between components, the diverse data fusion patterns, and the presence of \sout{branch-then-join nodes}\textcolor{blue}{branch-then-fusion paths} create a complex structure that demands a robust framework to manage real-time constraints, as delays in any component can compromise vehicle safety. 
Motivated by this, the real-time systems community has introduced several key metrics to ensure timely responses. 
\begin{itemize}
    \item \textbf{Maximum Reaction Time (MRT):} The longest time interval from the occurrence of an external event (which can happen between sensor instances) to the first actuator response to that event~\cite{durr2019end}.\footnote{A recent study has reported that MRT is equivalent to the maximum data age (MDA) \cite{gunzel2023equivalence}; hence, we use MRT as a representative of both in this work.} 
    For a DAG with multiple sensors contributing to an actuator, MRT should consider the maximum value across all sensor-to-actuator paths.
    \item \textbf{Maximum Time Disparity (MTD):} The maximum time difference between the release times of the oldest and newest sensor data contributing to an actuator output instance, measured across all instances of that actuator \cite{jiang2023analysis}.
    \item \textbf{Worst-Case Response Time (WCRT):} The maximum duration of each chain in a DAG, measured from a sensor task's release time to an actuator task's completion time~\cite{sobhani2023timing}.
    \item \textbf{Peak Age of Information (PAoI):} AoI represents the time elapsed since the generation of the latest data sample to the current time \cite{wang2022distributed}. PAoI is the maximum AoI value across all sensors in the system \cite{cao2020peak}.
\end{itemize}

\sout{Directed Acyclic Graphs (DAGs) have proven effective in representing the complex structure of ADS software stack and all dependencies among components, supporting the on-time coordination of all functionalities in real-time~\cite{xu2022aoi, sun2023real}.
Within a DAG representation, sensors and actuators can be represented as source and sink nodes, respectively, aiding in the analysis and bounding of these timing metrics by tracking the release times of source nodes and completion times of sink nodes. The following sections explore our detailed DAG modeling with diverse fusion types and discuss how these real-time performance metrics can be optimized by considering the effects of different fusion nodes, using ILP for quantitative comparison.}

\subsection{System Model}\label{sec:sys_model}

This section presents our system model, defining DAG, chain, and task representations on a multi-core platform with identical CPU cores. \sout{We use a discrete-time model, where time intervals are non-negative integer multiples of a defined system time unit (e.g., time tick).}\textcolor{blue}{The summary of notations is provided in Table~\ref{tab:notations}.}

\smallskip
\noindent\textbf{DAG.}
A Directed Acyclic Graph (DAG) is used to represent the ADS software application. We denote the DAG as $G=(V, E)$, where $V$ represents the set of vertices and $E$ represents the set of edges. Each vertex or node in the DAG corresponds to a task, and the terms ``vertex'', ``node'', and ``task'' are used interchangeably throughout the paper. Each edge $E_{i', i} \in E$ indicates a data dependency, where the output of task $i'$ serves as the input to task $i$. Without loss of generality, we assume inter-task communication occurs through a unit-sized buffer between consecutive tasks, following the ``last-is-best'' principle \cite{li2024priority}, which ensures that only the most recent data is retained. Additionally, the buffer is non-blocking \cite{abdullah2019worst}, which is easily achievable through double-buffering techniques within the implicit communication paradigm \cite{li2024priority}. Source and sink nodes of the DAG, respectively, are the representative of tasks from the sensor and actuator classes mentioned in Sec.~\ref{sec:dag_ads}. We assume the DAG under analysis, $G$, comprises $m$ sensor nodes and $n$ non-sensor nodes, with a total size of $|V| = m + n$.

\smallskip
\noindent\textbf{Chain.}
A chain represents a path within the DAG, running from a source node (sensor) to a sink node (actuator) and capturing a sequence of dependent tasks in a set order. Together, these chains form the complete DAG, showing all interdependencies in the system. Since our focus is on the overall DAG structure rather than individual chains, we use DAG notations.

\smallskip
\noindent\textbf{Tasks.}
A task $\tau_i$ is a software component represented as a node in the DAG (i.e., $V=\{ \tau_1, \dots, \tau_{m+n} \}$), which can be triggered either by a \textit{timer} or by an external \textit{event}. The task $\tau_i$ is characterized as \textcolor{blue}{$\tau_i = \{e_i, T_i, D_i, \theta_i, pred_i\}$; where $e_i$ is the worst-case execution time (WCET) of $\tau_i$; $T_i$ and $D_i$ denote its period and relative deadline, respectively; $\theta_i$ represents the type of $\tau_i$ (e.g., sensor, fusion, etc; details will be followed);} and $pred_i$ is the set of adjacent predecessor tasks of $\tau_i$ in the DAG, i.e., $pred_i=\{ \tau_{i'}| \forall E_{i', i} \in E\}$. Note that for an event-triggered task, we define $T_i = 0$. \sout{Without loss of generality, w}\textcolor{blue}{W}e assume that the execution of each task follows the \textit{read-execute-write} semantics, where input data is read at the task's start time, and output data is written at its finish time \cite{becker2016synthesizing, choi2020chain, teper2022end}. Tasks are scheduled non-preemptively, like \textit{callbacks} in ROS 2 and \textit{runnables} in AUTOSAR.

\smallskip
\noindent\textbf{Task Instances.}
Each task, whether triggered by a timer or an event, generates a sequence of task instances. We denote the $j$-th instance of task $\tau_i$ as $\tau_{i,j}$. The start and finish times of \textcolor{blue}{execution of} $\tau_{i,j}$ are represented by $s_{i,j}$ and $f_{i,j}$, respectively. For simplicity, we assume that a task instance is \textcolor{blue}{released (ready for execution) immediately upon being triggered, 
and its absolute deadline, denoted as $d_{i,j}$, is calculated by this release time $r_{i,j}$ plus its relative deadline $D_i$. Also, the first instance of each sensor task is assumed to arrive at time 0.} 

\sout{The number of instances of a task over a time interval of interest $\Delta$ varies based on the task type. A task type is determined by its triggering behavior and the number of preceding inputs, which will be discussed in the following section. The summary of notations is provided in Table~\ref{tab:notations}.}


\begin{table}[t]
\centering
\caption{Table of Notations}
\label{tab:notations}
\footnotesize
\begin{tabular}{|c|p{0.8\columnwidth}|}
\hline
\textbf{Symbol} & \textbf{Description} \\ \hline
$G = (V,E)$ & DAG $G$ with the set of vertices $V$ and edges $E$ \\ \hline
$\tau_i$ & The task $i$ where $\tau_i \in V$\\ \hline
\textcolor{blue}{$e_i, T_i, D_i$} & \textcolor{blue}{The execution time, the period, and the deadline of $\tau_i$} \\ \hline
$\theta_i$ & The type of $\tau_i$ where $\theta_i \in \{ \text{``sen'', ``sub'', ``t-fus'', ``w-fus'', ``i-fus''} \}$. \\ \hline
$pred_i$ & The set of adjacent predecessors of $\tau_i$ where $pred_i=\{ \tau_{i'}| \forall E_{i', i} \in E\}$ \\ \hline
$\tau_{i,j}$ & The $j$-th instance of $\tau_i$ \\ \hline
\textcolor{blue}{$s_{i,j}, f_{i,j}, d_{i,j}$} & \textcolor{blue}{The start time, finish time, and absolute deadline of instance $\tau_{i,j}$} \\ \hline
$m, n, \Pi$ & The number of sensors, non-sensor tasks, and cores \\ \hline
\textcolor{blue}{$\tau_s, \tau_{\otimes}$} & \textcolor{blue}{A source and a sink node in the DAG} \\ \hline
\end{tabular}
\end{table}

\section{Task Types and Fusion Patterns in ADS}\label{sec:model_fusion}
In this section, we characterize the diverse behaviors of various types of tasks in ADS, with a particular focus on fusion tasks. Based on various triggering factors and the number of precedent inputs, we categorize a task $\tau_i$ into one of five distinct types, which are detailed in the following paragraph.
\begin{itemize}
    \item Sensor ($\theta_i$ = ``sen''): Sensor task is a periodic timer-triggered task, sampling real-time data and generating output for a successor task. Considering a sensor task is a source node in DAG, it has no predecessor dependencies. Therefore, $pred_i = \varnothing$ is an empty set for a sensor task $\tau_i$.
    
    \item Subscription ($\theta_i$ =``sub''): Subscription task is event-triggered, activated upon receiving one input message and publishing one output message accordingly. For these tasks, the size of the $pred_i$ is always 1, meaning $\| pred_i \| =1$.
    
    \item T-fusion ($\theta_i$ = ``t-fus''): T-fusion task is \textbf{t}imer-triggered and periodic, gathering inputs from multiple preceding tasks and merging them into a unified output. \textcolor{blue}{A real-world ADS example of this type can be found in Autoware, such as the \texttt{ndt\_scan\_matcher} component~\cite{autoware}.} Notably, an intermediate or sink node in a DAG that is timer-triggered but has only one input can also be considered a special form of T-fusion.
    \item W-fusion ($\theta_i$ = ``w-fus''): W-fusion is a \textbf{w}ait-for-all fusion task, triggered when it has received inputs from all its predecessor tasks, only generating an output once all required inputs are available. \textcolor{blue}{A real-world ADS example of this type in Autoware is \texttt{tracking\_object\_merger} component~\cite{autoware}.}
    \item I-fusion ($\theta_i$ = ``i-fus''): In contrast to W-fusion, an I-fusion task is triggered as \textbf{i}mmediately as any of its predecessor tasks publishes a new message, reacting without waiting for all inputs to arrive. \textcolor{blue}{An example of this type in Autoware is the \texttt{traffic\_light\_occlusion\allowbreak\_predictor} component~\cite{autoware}.} 
\end{itemize}

Among these task types, sensor and subscription nodes have been \textcolor{blue}{extensively} studied in prior work. Their modeling and scheduling are relatively straightforward, and we include them here for the sake of completeness. \textcolor{blue}{However, unlike prior work where each study considered only one type of fusion node, our main focus is on all three fusion types described above. For example, Toba et al. \cite{toba2024deadline} considered \textit{trigger} and \textit{update} edges, where only \textit{trigger} edges activate the fusion node’s execution. Similarly, Teper et al. \cite{teper2022end} used \textit{passive} and \textit{trigger} edges, with the fusion node triggered solely by the one dominant trigger edge. These behaviors fall into our W-fusion category, which can also handle cases where the triggering edge is not predetermined. Multi-rate DAG analysis \cite{sun2023real, li2024priority} assumes that all fusion nodes are time-triggered, which can be represented using our T-fusion nodes. Sun et al. \cite{sun2025jointly} assumed that only sensors and actuators are timer-triggered and all other intermediate nodes, including fusion nodes, are event-triggered, which our I-fusion node can model.}

To better illustrate these task types, Fig.~\ref{fig:dag_types} presents an example DAG with the node types defined above. Nodes $\tau_1$ to $\tau_4$ represent sensors, $\tau_5$ and $\tau_9$ are subscription nodes, and $\tau_8$, $\tau_{10}$, and $\tau_{11}$ correspond to I-fusion, T-fusion, and W-fusion nodes, respectively. Our model supports not only multiple sink nodes but also branch (or fork) nodes. A branch node whose output diverges into multiple paths can be any of the above node types.  
If the immediate successor of a branch node is a subscription node, e.g., $\tau_6$ and $\tau_7$, it is modeled as ``i-fus'' for formulation purposes \textcolor{blue}{(see Appendix~\ref{sec:branch} for details)}. 

Based on task types, the number of instances of each task \sout{over a given time interval $\Delta$ is calculated differently, assuming that all arriving instances have their deadlines within this interval}\textcolor{blue}{can be calculated. Consider the time interval of interest $\Delta$ that is an integer multiple of the hyperperiod (HP), i.e., $\Delta = k\cdot HP$ where $HP$ is the least common multiple (LCM) of all time-triggered tasks (sensors and T-fusion nodes). The interval $\Delta$ starts at time 0, and all sensor tasks initially arrive at time 0.}
Alg.~\ref{al:n-ins} details the calculation of the instance count, $\textit{n-ins}$, for a task $\tau_i$ within this interval \textcolor{blue}{$\Delta$}. For timer-triggered tasks like sensor and T-fusion nodes, the instance count \sout{follows conventional periodic task behavior, given by $\textit{n-ins}(\tau_i, \Delta) = \left\lceil \frac{\Delta}{T_i} \right\rceil$}\textcolor{blue}{is straightforward to compute by $\textit{n-ins}(\tau_i, \Delta) = \frac{\Delta}{T_i}$ because $\Delta$ is an integer multiple of HP} (lines 2-3). For event-triggered tasks, the instance count calculation varies. The instance count of a W-fusion node is determined by the least frequent adjacent predecessor’s $\textit{n-ins}$ (lines 4-5), \textcolor{blue}{as the W-fusion has to wait for all predecessors by its definition. An I-fusion node is triggered by any updated input from its predecessors. But at the very beginning, it still has to wait for every predecessor to produce at least one input for fusion. Hence, the instance count of an I-fusion node (lines 6-7) is obtained by the sum of instance counts of its predecessors ($\sum \textit{n-ins}(pre, \Delta)$) subtracted by the number of predecessor instances to wait ($\|pred_i \|-1$).}
In the case of a subscription task, the instance count matches the $\textit{n-ins}$ of its adjacent predecessor, \sout{which could be either a timer-triggered or event-triggered node}\textcolor{blue}{as it has a single predecessor and is triggered whenever input is received from its predecessor.} (lines 8-9).

\vspace{-5mm}
\noindent\begin{minipage}[!t]{.4\textwidth}
\begin{figure}[H]
\includegraphics[width=.9\linewidth, keepaspectratio]{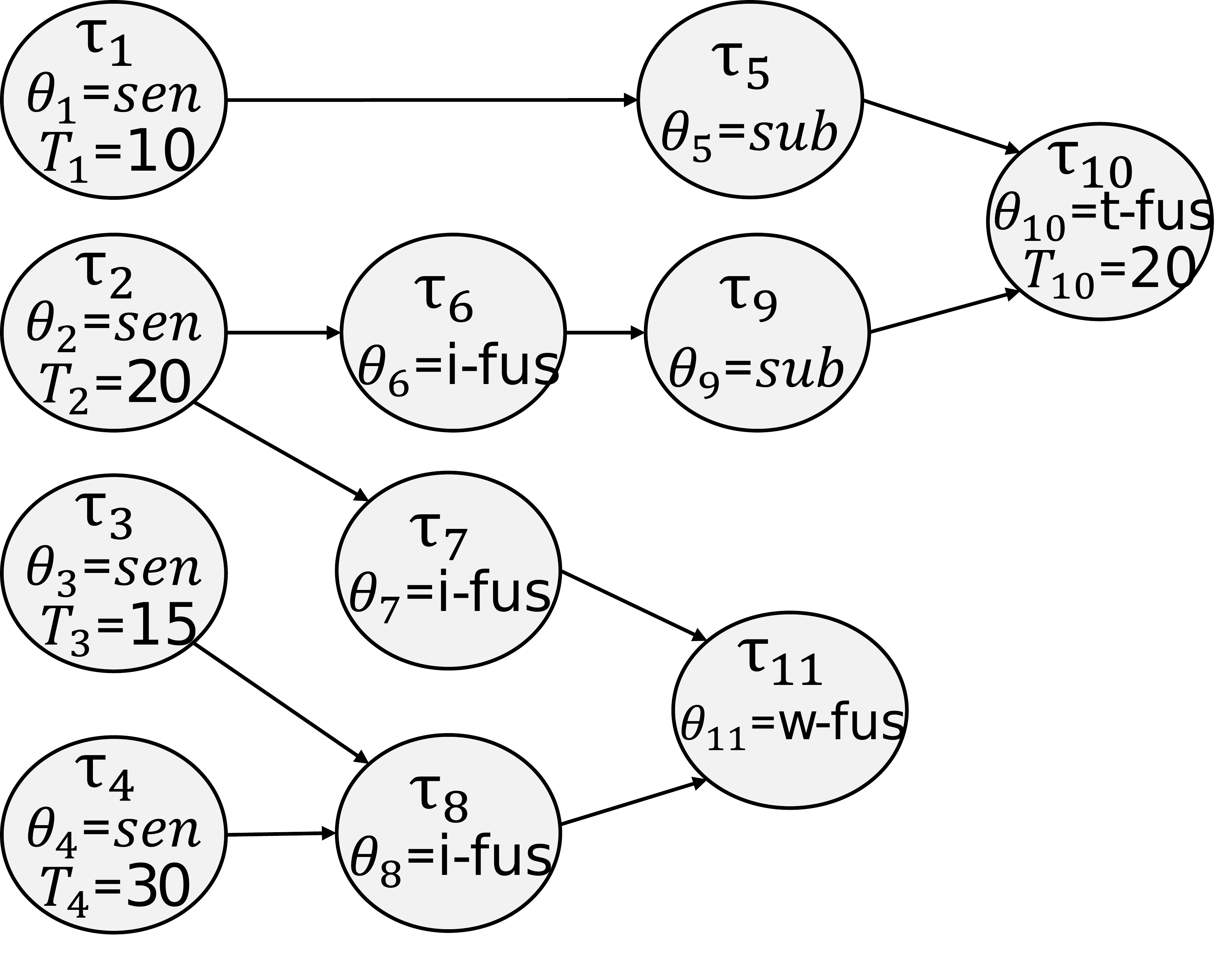}
\caption{Task types in a DAG}
\label{fig:dag_types}
\end{figure}
\end{minipage}%
\begin{minipage}[!t]{.6\textwidth}
\begin{algorithm}[H]
\caption{n-ins($\tau_i$,$\Delta$)}\label{al:n-ins}
    \begin{algorithmic}[1]
    \State $\textit{n-ins} \leftarrow 0$
    \If{$\theta_i \in \{ \text{``sen''}, \text{``t-fus''}\}$}
    \State $\textit{n-ins}\leftarrow$ \sout{$\lceil \frac{\Delta}{T_i} \rceil$}
    \textcolor{blue}{$\frac{\Delta}{T_i}$}
    \ElsIf{$\theta_i == \text{``w-fus''}$}
    \State \textcolor{blue}{$\textit{n-ins}\leftarrow \min_{pre \in pred_i} \big ( \textit{n-ins}(pre, \Delta) \big )$}
    \ElsIf{$\theta_i == \text{``i-fus''}$}
    \State $\textit{n-ins}\leftarrow$
    \Statex \quad\quad$\big (\sum_{pre \in pred_i} \textit{n-ins}(pre, \Delta) \big ) - (\|pred_i \|-1)$
    \ElsIf{$\theta_i == \text{``sub''}$}
    \State \textcolor{blue}{$\textit{n-ins}\leftarrow \textit{n-ins}(pred_i, \Delta)$}    
    \EndIf \hspace{2mm}
    \Return $\textit{n-ins}$
    \end{algorithmic}
\end{algorithm}
\end{minipage}%
\vspace{5mm}

\noindent\textbf{\textcolor{blue}{Example: Instance Counts.}}
\textcolor{blue}{
Let us apply Alg.~\ref{al:n-ins} to the DAG example in Fig.~\ref{fig:dag_types}, assuming $\Delta=1\cdot HP = 60$. For the sensor tasks $\tau_1, \tau_2, \tau_3, \tau_4$, Alg.~\ref{al:n-ins} gives $\textit{n-ins}(\tau_1, HP) =6, \textit{n-ins}(\tau_2, HP) =3, \textit{n-ins}(\tau_3, HP) =4, \textit{n-ins}(\tau_4, HP) =2$.
For the subscription task $\tau_5$, it inherits $\tau_1$, so $\textit{n-ins}(\tau_5, HP)=\textit{n-ins}(\tau_1, HP)=6$. 
For the I-fusion tasks with single predecessors, $\tau_6$ and $\tau_7$, Alg.~\ref{al:n-ins} gives $\textit{n-ins}(\tau_6, HP)=\textit{n-ins}(\tau_2, HP) - (1-1) = 3$ and $\textit{n-ins}(\tau_7, HP)=\textit{n-ins}(\tau_2, HP) - (1-1) = 3$, showing that they are triggered by every instance of their predecessors. 
For the other I-fusion task $\tau_8$, $\textit{n-ins}(\tau_8, HP)=\textit{n-ins}(\tau_3, HP) + \textit{n-ins}(\tau_4, HP)- (2-1) = 5$ because the first instance $\tau_{8,1}$ is triggered only after both $\tau_{3,1}$ and $\tau_{4,1}$ have arrived. 
For the subscription task $\tau_9$, $\textit{n-ins}(\tau_9, HP)=\textit{n-ins}(\tau_6, HP)=3$. For the T-fusion task $\tau_{10}$, $\textit{n-ins}(\tau_{10}, HP)=3$. For the W-fusion task $\tau_{11}$, $\textit{n-ins}(\tau_{11}, HP)=min(\textit{n-ins}(\tau_7, HP), \textit{n-ins}(\tau_8, HP)) =  min(3, 5) = 3$ because $\tau_{11}$ waits for new inputs from both $\tau_7$ and $\tau_8$ before triggering a new instance.}

\begin{figure}[t]
    \centering
    \begin{minipage}[c]{0.49\linewidth}
        \centering
        \subfloat[DAG structure]{\includegraphics[width=0.8\linewidth, height=2cm]{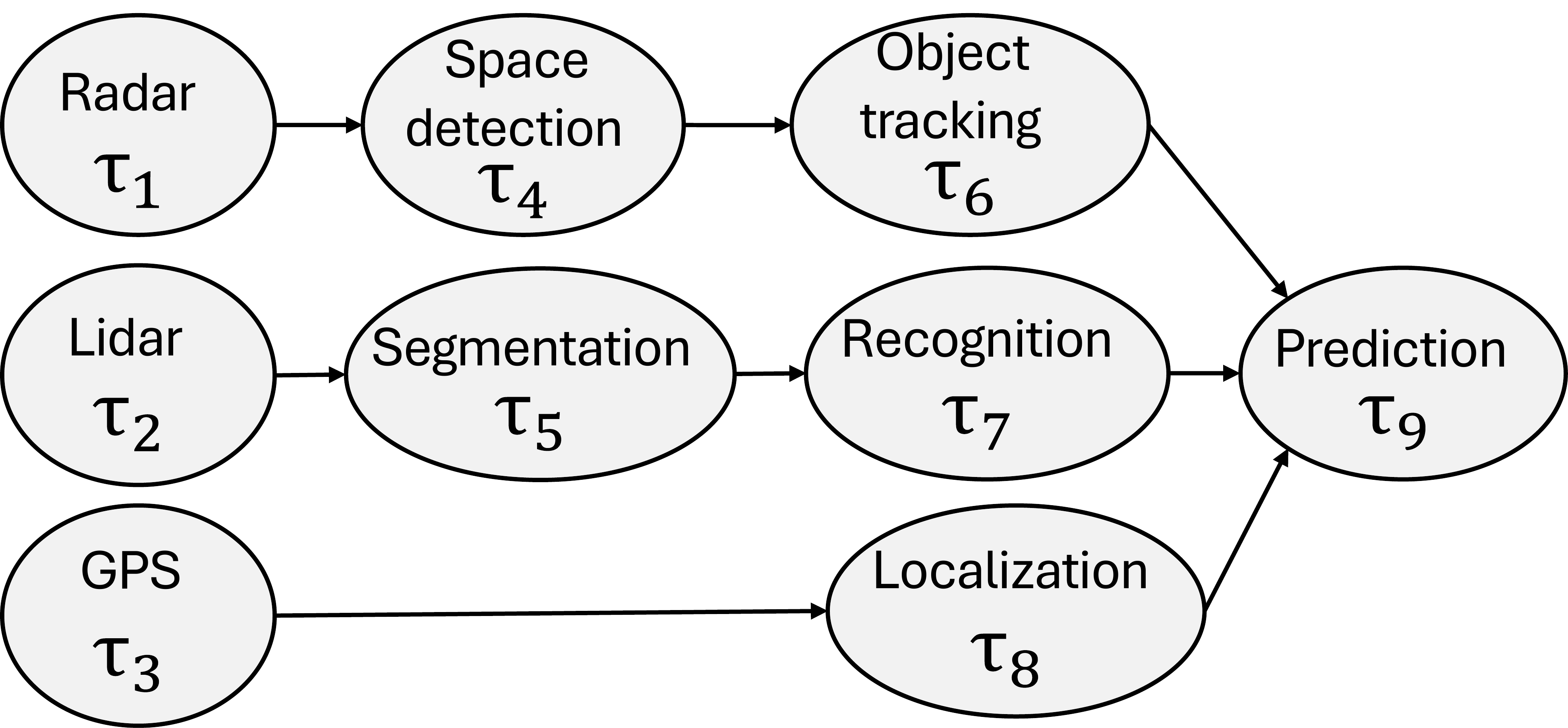}\label{fig:simple_dag}\vspace{2mm}} \\
        \subfloat[DAG scheduling with T-fusion node]{\includegraphics[width=\linewidth]{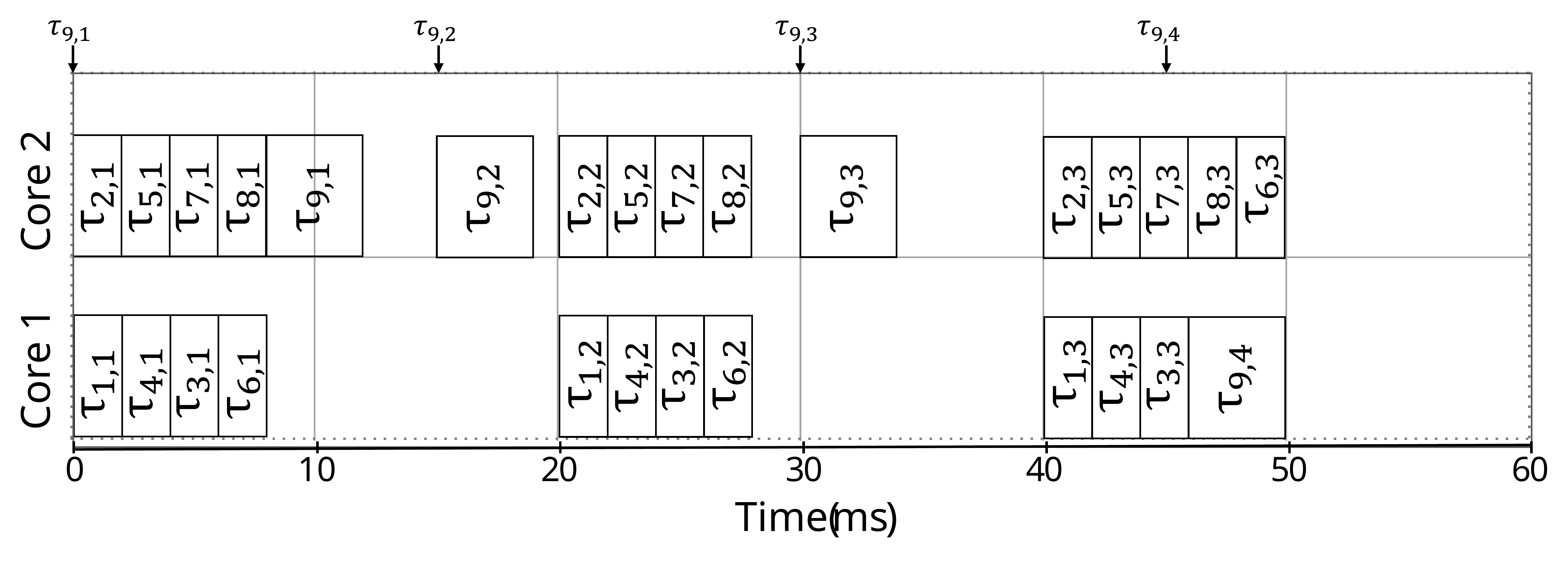}\label{fig:simple_tfus}}
    \end{minipage}
    \begin{minipage}[c]{0.49\linewidth}
        \centering
        \subfloat[DAG scheduling with W-fusion node]{\includegraphics[width=\linewidth]{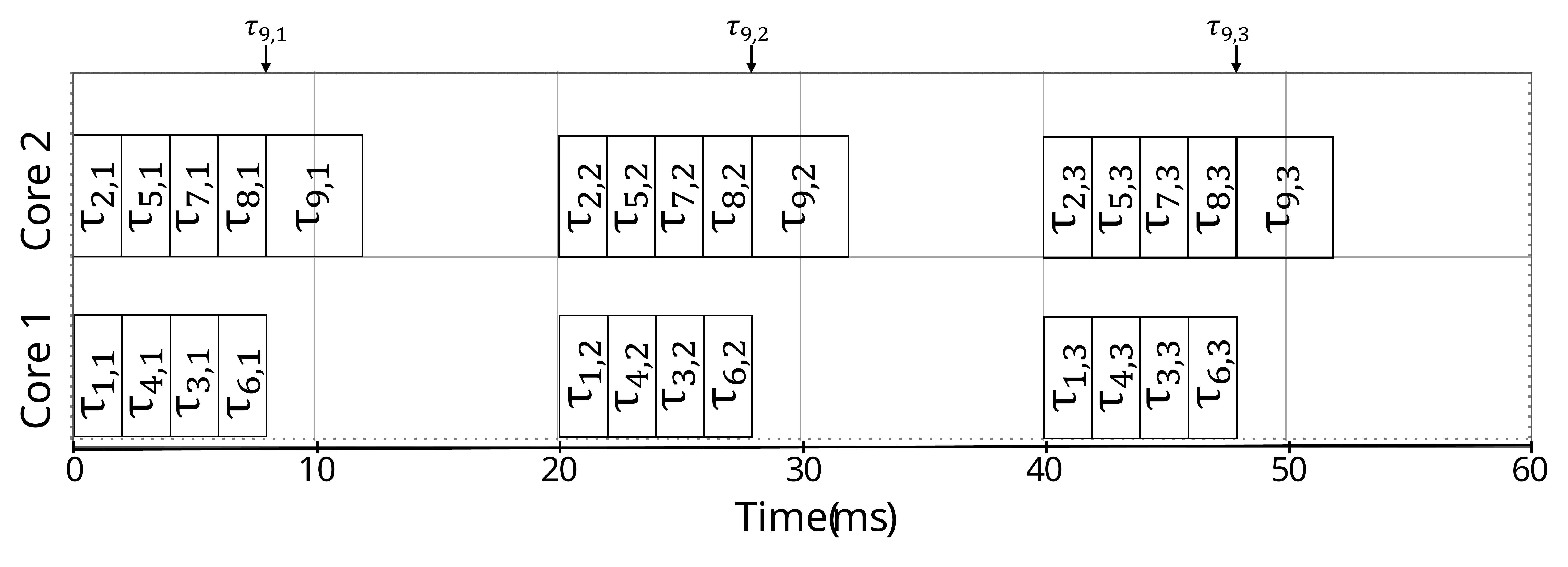}\label{fig:simple_wfus}} \\
        \subfloat[DAG scheduling with I-fusion node]{\includegraphics[width=\linewidth]{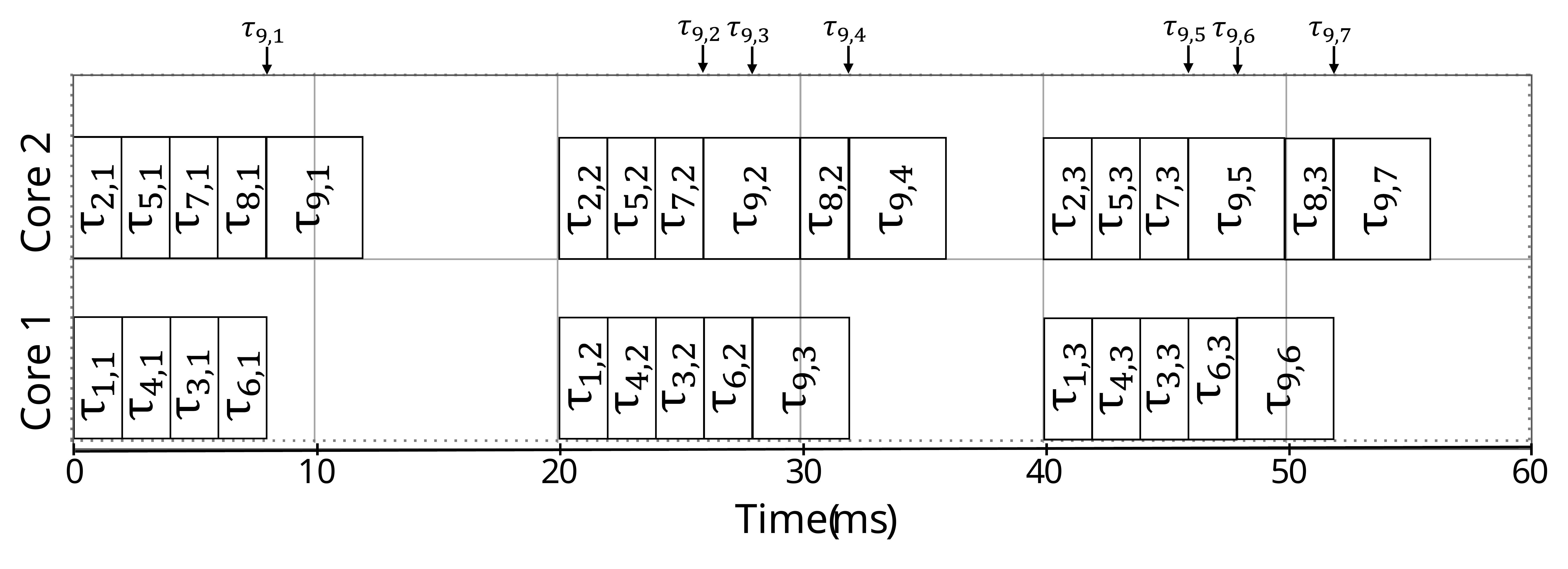}\label{fig:simple_ifus}}
    \end{minipage}
    \caption{Illustrative examples on fusion nodes}
    \label{fig:fusions_illustrative}
\end{figure}

\noindent\textbf{\textcolor{blue}{Example: Scheduling Behavior.}}
To clarify the behavior of fusion nodes in the scheduling context, we provide illustrative examples for the three fusion types, shown in Fig.~\ref{fig:simple_dag}\sout{.
This }\textcolor{blue}{, which} is a subset of the ADS software stack in Fig.~\ref{fig:ads}\sout{, representing the data flow from \textit{radar}, \textit{lidar}, and \textit{gps} sensors to the \textit{prediction} node (Fig.~\ref{fig:simple_dag})}. In this DAG, the \textit{prediction} node serves as a fusion node, while all other non-sensor tasks are subscription nodes. We set the WCET of sensor and subscription nodes to 2 ms and assigned a period of 20 ms to each sensor. The WCET of the fusion node was set to 4 ms, and we explored its behavior by varying its type among T-fusion, W-fusion, and I-fusion. We first configure the \textit{prediction} task as a T-fusion node with a period of 15 ms and examine its effect on DAG scheduling, as shown in Fig.~\ref{fig:simple_tfus}. In this case, the \textit{prediction} task is triggered once per its period ($=15$), regardless of input arrivals. Next, to illustrate W-fusion, we change the \textit{prediction} task to have the type $\theta = \text{``w-fus''}$. The scheduling behavior of W-fusion, waiting for all its dependencies before execution, is depicted in Fig.~\ref{fig:simple_wfus}. Finally, we demonstrate I-fusion by setting the \textit{prediction} task to have $\theta = \text{``i-fus''}$. As shown in Fig.~\ref{fig:simple_ifus}, I-fusion triggers the \textit{prediction} task immediately upon the arrival of any input, \textcolor{blue}{except the first instance,} highlighting its responsiveness compared to the other fusion types. \textcolor{blue}{The first instance of each fusion node requires all its predecessors to be executed at least once. The arrows in the figures indicate the triggering time of each fusion instance.}

Now that we have a clearer understanding of fusion tasks and their impact on scheduling and real-time performance, we formulate their diverse behaviors and analyze their effects on key metrics such as MRT, MTD, PAOI, and WCRT in the following section.

\section{Formulation of Fusion Tasks\sout{ and Their Impact on Real-Time Performance Metrics}}\label{sec:formulation}
In this section, we begin by providing a brief formulation of the general requirements for resource allocation and task scheduling in a multi-core system, using ILP constraints. We then focus on task-specific behaviors, particularly fusion tasks, and explore how these tasks influence scheduling and timing, especially their impact on the start and finish times of other tasks. Finally, we explain how to model and formulate key real-time performance metrics for ADS into ILP, enabling the study of the effects of different fusion types on these metrics.

The DAG under analysis, $G$, operates on a multi-core platform equipped with $\Pi$ identical CPU cores. \sout{As tasks can be globally scheduled across cores, meaning that instances of a task may execute on different cores.} \textcolor{blue}{We do not restrict tasks to be bound to specific cores, allowing instances of the same task to execute on different cores. To achieve this,} we define a binary variable $y^{\pi}_{i,j}$ \sout{to indicate whether}\textcolor{blue}{that equals 1 if} an instance $\tau_{i,j}$ is assigned to core $\pi$ (where $1 \leq \pi \leq \Pi$)\textcolor{blue}{, and 0 otherwise}. In a scheduling problem, once a task instance is assigned to a core, its start and finish times need to be determined. Since tasks are non-preemptible, each task runs uninterrupted once it starts. Therefore, given the start time, we can calculate the finish time by adding the WCET $e_i$ of the task. As a result, the scheduling problem reduces to determining the start time of each task instance\textcolor{blue}{; once the start time is known, the finish time follows directly. To ensure each task instance $\tau_{i,j}$ meets its deadline, we compare its finish time $f_{i,j}$ with its absolute deadline $d_{i,j}$.}
Since expressing basic scheduling requirements, \textcolor{blue}{such as task-to-core mapping, overlap prevention, task deadline constraints ($D_{i}$ and $d_{i,j}$),} in ILP is well-known and not the main focus of our work, we present them in the supplemental material (Appendix~\ref{sec:general}).

Before focusing on how to formulate the different fusion types into ILP constraints, we briefly review sensor and subscription nodes for the sake of completeness. For sensor nodes, which have no preceding data dependencies, the start time is determined solely by their timer period and instance release time, following the standard definition of periodic tasks. To ensure that a sensor instance $\tau_{i,j}$ (where $\theta_i = \text{``sen''}$) starts after its release time and finishes before the next instance arrives, we define the following constraints:
\begin{equation}\label{eq:sen_start_ge_release}
\begin{split}
s_{i,j} &\geq T_i \cdot (j-1), \quad \forall \tau_i \in V \textcolor{blue}{\land ~ \theta_i=\text{``sen''}}, \forall j \in [1, \textit{n-ins}(\tau_{i}, \Delta)]\\
s_{i,j} &\geq f_{i,j-1}, \quad \quad \quad \forall \tau_i \in V \textcolor{blue}{\land ~ \theta_i=\text{``sen''}}, \forall j \in [1, \textit{n-ins}(\tau_{i}, \Delta)]    
\end{split}
\end{equation}

For subscription nodes, the start time is determined by their dependency on a single predecessor node. Specifically, the start time of a subscription task instance must follow the finish time of its preceding task instance with the same instance index. Therefore, for a subscription task instance $\tau_{i,j}$ (where $\theta_i = \text{``sub''}$), the following equation holds, with $\tau_{i'}$ representing its sole predecessor:
\begin{equation}
s_{i,j} \geq f_{i',j}, \quad \textcolor{blue}{\forall \tau_i \in V \land ~ \theta_i=\text{``sub''},~} pred_i = [\tau_{i'}], \forall j \in [1, \textit{n-ins}(\tau_{i}, \Delta)]    
\end{equation}

\subsection{Fusion Task Constraints}\label{sec:formulate_fusion}
Unlike subscription tasks, fusion tasks (like T-fusion, W-fusion, or I-fusion) have multiple predecessors with different arrival rates. Hence, it is not straightforward to establish the relationship between the instance index of a fusion task and those of its predecessors. 

To address this issue, we first introduce the concept of \textit{producer} tasks. Producer tasks are sensor and fusion nodes; they generate and provide data to subsequent tasks, establishing the root for a new instance for subsequent \textcolor{blue}{non-producer} tasks. Hence, any non-producer tasks (i.e., subscription tasks) use the same instance indices as those of their producers' instances. 

Let us use $producer(\tau_i)$ to denote the producer task of $\tau_i$. If $\tau_i$ is a subscription task, $producer(\tau_i)$ obviously gives only one task: the closest preceding sensor or fusion task of $\tau_i$ in the DAG hierarchy. If $\tau_i$ is either a sensor or fusion task, $producer(\tau_i)=\tau_i $ because $\tau_i$ is a producer by itself. Note that $producer(\tau_i)$ should not be confused with $pred_i$: if $\tau_i$ is a fusion task, it can have multiple predecessors ($|pred_i| \geq 1$) and each of its predecessors has its own producer ($producer(\tau_{i'})| \tau_{i'}\in pred_i$). \textcolor{blue}{We also define 
$pred\_producers(\tau_i)$ as the set of producers of $\tau_i$'s predecessors:
$$pred\_producers(\tau_i) = \{producer(\tau_{i'})| \forall \tau_{i'} \in pred_i\}$$
For example, consider the DAG in Fig.~\ref{fig:simple_dag}. The sink node $\tau_9$ has $pred_9=\{ \tau_6, \tau_7, \tau_8\}$ and $pred\_producers(\tau_9)= \{ \tau_1, \tau_2, \tau_3\}$. Note that $producer(\tau_9) = \tau_9$ because $\tau_9$ is a fusion node. 
}

Next, to represent which \textcolor{blue}{instances of predecessors contribute} to a fusion task instance, we define a binary decision variable as follows:


\begin{definition}\label{def:u}
\textcolor{blue}{Consider a fusion task $\tau_i$ and one of its predecessors' producer $\tau_p$ where $ \tau_p\in pred\_producers(\tau_i)$.
A binary decision variable $u_{(i,j),(p,j_p)}$ indicates whether the fusion task's instance $\tau_{i,j}$ uses the $j_p$-th instance of $\tau_p$:}
\begin{equation}\label{eq:u_def}
u_{(i,j),(p,j_p)}= \begin{cases}
1, & \scalebox{1}{\text{\small{if $\tau_{i,j}$ uses the $j_p$-th instance of $\tau_p$}}} \\
0, & \text{otherwise}
\end{cases}
\end{equation}
\end{definition}

\sout{With this definition, we can now set constraints on the start time of a fusion task's instance by linking it to the finish time of its predecessor's instance (identified by $u_{(i,j),(i',j')}$). This approach ensures that each fusion instance begins only after the appropriate predecessor instance completes.}
\textcolor{blue}{With this definition, for a fusion task instance $\tau_{i,j}$, we can determine which instance $j'$ of $producer(\tau_{i'}\in pred_i)$ is used by $\tau_{i,j}$. 
Recall that $producer(\tau_{i'})$ determines the instance index of the subsequent task $\tau_{i'}$; this applies to any type of task, because if $\tau_i'$ is either a sensor or fusion task, $producer(\tau_{i'}) = \tau_i'$.
Therefore, we can set constraints on the start time of the fusion task instance $\tau_{i,j}$ by linking it to the finish time of the predecessor instance $\tau_{i',j'}$ it uses.
This ensures that each fusion instance begins only after the appropriate predecessor instance completes.}


Hence, we establish the following lemma for all three types of fusion nodes.

\begin{lemma}[Constraint: start time of fusion nodes]\label{lem:start_fusion}
The start time of a fusion task instance $\tau_{i,j}$, i.e., $\theta_i \in \{\text{``t-fus'', ``w-fus'', ``i-fus''}\}$, should satisfy the following condition:
\begin{equation}
s_{i,j} \geq f_{i',j'} \cdot u_{(i,j),(producer(\tau_{i'}), j')}, \text{ \scalebox{1}{$\forall \tau_{i'} \in pred_i, \forall j' \in [1, \textit{n-ins}(\tau_{i'}, \Delta)]$}}    
\end{equation}
\begin{proof}
Here, $\tau_{i'}$ is one of $\tau_i$'s predecessors ($\forall \tau_{i'} \in pred_i$), and $u_{(i,j),(producer(\tau_{i'}), j')}$ indicates whether $\tau_{i,j}$ uses $j'$-th instance of $\tau_{i'}$'s producer. 
To prove this equation, we consider two cases for $u_{(i,j),(producer(\tau_{i'}), j')}$:
\textit{(i)} If $u_{(i,j),(producer(\tau_{i'}), j')} = 0$, the right-hand side is zero, which is always valid as start times are never negative. \textit{(ii)} If $u_{(i,j),(producer(\tau_{i'}), j')} = 1$, meaning the $j'$-th instance of $\tau_{i'}$'s producer is used by $\tau_{i,j}$, we have two sub-cases:

\noindent \textit{(ii-A)} If $\tau_{i'}$ is a sensor or fusion \textcolor{blue}{node}, $producer(\tau_{i'}) = \tau_{i'}$, and $u_{(i,j),(producer(\tau_{i'}), j')} = u_{(i,j),(i', j')} = 1$. This means \sout{the $j'$-th instance of $\tau_{i'}$}\textcolor{blue}{$\tau_{i',j'}$} contributes to $\tau_{i,j}$, so $s_{i,j} \geq f_{i',j'}$.

\noindent \textit{(ii-B)} If $\tau_{i'}$ is a subscription \textcolor{blue}{node}, it inherits the instance index of its producer. Thus, the $j'$-th instance of $\tau_{i'}$ is used by $\tau_{i,j}$, and $s_{i,j} \geq f_{i',j'}$ holds.\qed
\end{proof}
\end{lemma}

Each instance of a fusion task \textcolor{blue}{uses} only one instance from each of its \textcolor{blue}{predecessors'} producers, specifically the most recent instance available from each producer. This imposes the following constraints.

\begin{lemma}[Constraint: used only once by an instance]\label{lem:u_only_once}
Since a fusion task instance $\tau_{i,j}$ uses only one instance from each of its predecessors' producers, the following constraint must hold for any $\tau_i \in V$, where $\theta_i \in \{\text{``t-fus'', ``w-fus'', ``i-fus''}\}$, and any $j \in [1, \textit{n-ins}(\tau_{i}, \Delta)]$: 
\begin{equation}
\sum_{j'=1}^{\textit{n-ins}(\tau_{i'}, \Delta)} u_{(i,j),(producer(\tau_{i'}), j')} = 1, \text{ \scalebox{1}{$\forall \tau_{i'} \in pred_i$}}    
\end{equation}
\begin{proof}
We prove by contradiction. Assume for a fusion instance $\tau_{i,j}$ and its predecessor $\tau_{i'}$, the summation can be other than 1. We consider two cases: 

\noindent \textit{(i)} $\sum_{j'=1}^{\textit{n-ins}(\tau_{i'}, \Delta)} u_{(i,j),(producer(\tau_{i'}), j')} = 0$: This means $u_{(i,j),(producer(\tau_{i'}), j')} = 0$ for all $j'$, meaning no instance of $producer(\tau_{i'})$ or $\tau_{i'}$ contributes to $\tau_{i,j}$. This contradicts our system model, which requires that at least one of the predecessors must execute to perform a fusion operation.

\noindent \textit{(ii)} $\sum_{j'=1}^{\textit{n-ins}(\tau_{i'}, \Delta)} u_{(i,j),(producer(\tau_{i'}), j')} > 1$: If the summation is greater than 1, multiple instances of $\tau_{i'}$ contribute to a single fusion instance $\tau_{i,j}$. This contradicts fusion node requirements, as each fusion node uses only the most recent instance from each predecessor. While some instances may be dropped (T-fusion or W-fusion) or reused (T-fusion or I-fusion), only one instance of $\tau_{i'}$ should contribute to $\tau_{i,j}$.

Thus, by contradiction, the summation must be 1, proving the lemma.\qed
\end{proof}
\end{lemma}

\begin{lemma}[Constraint: most recent instance is used]\label{lem:u_most_recent}
Since a fusion task instance $\tau_{i,j}$ uses the most recent instance from each of its predecessor tasks' producers, the following constraint must hold for any $\tau_i \in V$, where $\theta_i \in \{\text{``t-fus'', ``w-fus'', ``i-fus''}\}$, and any $j \in [1, \textit{n-ins}(\tau_{i}, \Delta)]$: 
\begin{equation}\begin{split}
&\sum_{j'=1}^{\textit{n-ins}(\tau_{i'}, \Delta)} j' \cdot u_{(i,j),(producer(\tau_{i'}), j')} \leq 
\sum_{j'=1}^{\textit{n-ins}(\tau_{i'}, \Delta)} j' \cdot 
u_{(i,j+1),(producer(\tau_{i'}), j')}, \text{ \scalebox{0.8}{$\forall \tau_{i'} \in pred_i$}}
\end{split}
\end{equation}
\begin{proof}
By Lemma~\ref{lem:u_only_once}, for any fusion instance $\tau_{i,j}$ and predecessor $\tau_{i'}$, only one instance $j'$ in $[1, \textit{n-ins}(\tau_{i'}, \Delta)]$ has $u_{(i,j),(producer(\tau_{i'}), j')} = 1$, with all other terms being 0. Multiplying each term by $j'$, we get $j' \cdot u_{(i,j),(producer(\tau_{i'}), j')} = j'$ for the contributing instance and 0 for others. Summing over all instances gives the instance index $j'$ of the single contributing instance, which represents the instance of $\tau_{i'}$ used in $\tau_{i,j}$. The inequality ensures that the instance of $\tau_{i'}$ used in $\tau_{i,j+1}$ is the same or newer than in $\tau_{i,j}$.\qed
\end{proof}
\end{lemma}

Having established the constraints for all fusion node types, we now outline the specific constraints for each type. 

\vspace{10pt}
\noindent\textbf{T-fusion}:
For T-fusion tasks, which are timer-triggered, \textcolor{blue}{E}quation (\ref{eq:sen_start_ge_release}) applies, along with the fusion-specific lemmas mentioned above.

\vspace{10pt}
\noindent\textbf{W-fusion}:
For W-fusion tasks, where an instance is triggered only after receiving inputs from all its predecessor tasks, an instance of each predecessor's producer can be used at most once by the fusion task instances. This leads to the following constraint.

\begin{lemma}[Constraint: used once by ``w-fus'']\label{lem:wfus_only}
Since, for a fusion task $\tau_i$, an instance of each predecessor's producer can be used at most once by the fusion task instances, the following constraint must hold for any $\tau_i \in V$ with $\theta_i=\text{``w-fus''}$: 
\begin{equation}
\sum_{j=1}^{\textit{n-ins}(\tau_{i}, \Delta)} u_{(i,j),(producer(\tau_{i'}), j')} \leq 1, \text{ \scalebox{0.7}{$\forall \tau_{i'} \in pred_i, \forall j' \in [1, \textit{n-ins}(\tau_{i'}, \Delta)]$}}
\end{equation}
\begin{proof}
We prove by contradiction. Assume for a W-fusion task $\tau_i$ and its predecessor $\tau_{i',j'}$, the summation value is greater than 1. This implies at least two instances of $\tau_i$ use the same $j'$-th instance of $\tau_{i'}$, which contradicts the W-fusion property. A W-fusion waits for all predecessor inputs to arrive, and once $\tau_{i',j'}$ is used, the next instance of $\tau_i$ must wait for a new instance of $\tau_{i'}$, not reuse $\tau_{i',j'}$.
\end{proof}
\end{lemma}

Unlike W-fusion, T-fusion and I-fusion tasks can use the same producer instance multiple times across different instances of the fusion task.

\vspace{10pt}
\noindent\textbf{I-fusion}:
In the case of I-fusion tasks, an instance of the fusion node is triggered by the arrival of \emph{any} instance of its predecessor tasks. The key distinction here is that, for I-fusion, at least one of the predecessor tasks must provide a new instance to trigger an I-fusion task instance. \sout{For the first instance of an I-fusion task, all input data from its predecessor tasks must be available. However, for subsequent instances, a new instance from any predecessor can trigger a new I-fusion task instance, allowing the I-fusion task to process inputs independently as they arrive. The following lemma ensures that at least one predecessor task provides a new instance, thereby triggering a new instance of the I-fusion task.}

\begin{lemma}[Constraint: at least one new instance for ``i-fus''] \label{lem:u_at_least_new}
The following constraint ensures that the I-fusion task $\tau_i$ progresses from instance $\tau_{i,j}$ to $\tau_{i,j+1}$, only when at least one of its predecessors' producer has triggered a new instance; where $\tau_i \in V$ with $\theta_i=\text{``i-fus''}$ and $j \in [1, \textit{n-ins}(\tau_{i}, \Delta)]$: 
\begin{equation}
\resizebox{\textwidth}{!}{$
\begin{split}
&\sum_{\tau_{i'} \in pred_i} \big ( \sum_{j'=1}^{\textit{n-ins}(\tau_{i'}, \Delta)} j' \cdot u_{(i,j+1),(producer(\tau_{i'}), j')} - 
\sum_{j'=1}^{\textit{n-ins}(\tau_{i'}, \Delta)} j' \cdot 
u_{(i,j),(producer(\tau_{i'}), j')} \big ) \geq 1
\end{split}
$}
\end{equation}
\begin{proof}
For a fusion instance $\tau_{i,j}$ and its predecessor $\tau_{i'}$, the term $\sum_{j'=1}^{\textit{n-ins}(\tau_{i'}, \Delta)} j' \cdot u_{(i,j),(producer(\tau_{i'}), j')}$ gives the instance index of $\tau_{i'}$ used in $\tau_{i,j}$. Similarly, for $\tau_{i,j+1}$, the expression $\sum_{j'=1}^{\textit{n-ins}(\tau_{i'}, \Delta)} j' \cdot u_{(i,j+1),(producer(\tau_{i'}), j')}$ gives the instance index of $\tau_{i'}$ used in that fusion instance. According to Lemma~\ref{lem:u_most_recent}, the difference between these two sums must be greater than or equal to 0. If the difference is 0, the same instance of $\tau_{i'}$ is used in both $\tau_{i,j}$ and $\tau_{i,j+1}$. If the difference is positive, a newer instance is used in $\tau_{i,j+1}$. To ensure that at least one predecessor triggers the I-fusion node with a newer instance, we sum these differences across all predecessors and confirm that the total is non-zero.\qed
\end{proof}
\end{lemma}

\subsection{Real-Time Performance Metric Constraints}\label{sec:goals}
Based on the formulation of fusion task behavior in Sec.~\ref{sec:formulate_fusion}, 
this section explains how to model and derive key real-time performance metrics—such as \sout{maximum reaction time (MRT), maximum time disparity (MTD), peak age of information (PAoI), worst-case response time (WCRT), and makespan (MS)}\textcolor{blue}{MRT, MTD, PAoI, and MS}—using our proposed ILP optimization method. These metrics are essential for assessing system responsiveness and ensuring optimized performance.
As they reflect the timing relationships between source (sensor\textcolor{blue}{, $\tau_s$}) and sink (actuator\textcolor{blue}{, $\tau_\otimes$}) nodes in the DAG, we must identify which instances of sensor nodes are linked to each instance $j$ of a sink node $\tau_\otimes$, i.e., $\tau_{\otimes, j}$, when there is a directed path from \textcolor{blue}{a sensor $\tau_s$} to $\tau_\otimes$.


To determine which instances of sensor nodes contribute to each fusion or sink node instances, we define an additional binary variable \textcolor{blue}{$U_{(i,j),(s, j_s)}$}, where $\tau_{i,j}$ is the $j$-th instance of task $\tau_i$ (either a sink or fusion node) and \textcolor{blue}{$\tau_{s,j_s}$ is the $j_s$-th instance of a sensor task $\tau_s$}. Specifically:
\begin{equation}\label{eq:bigU_def}\textcolor{blue}{
U_{(i,j),(s,j_s)}= \begin{cases}
1, & \scalebox{1}{\text{\small{if instance $\tau_{i,j}$ uses the $j_s$-th instance of sensor $\tau_s$}}} \\
0, & \text{otherwise}
\end{cases}
}
\end{equation}
Note that $U$ differs from the decision variable $u$ (Def.~\ref{def:u}). While \textcolor{blue}{$u_{(i,j),(p, j_p)}$} represents direct relationships between fusion instances and their immediate predecessors' producers (determined by the ILP solver), \textcolor{blue}{$U_{(i,j),(s, j_s)}$} captures the end-to-end relationships between any task instance and sensor instances (computed using the $u$ variables).
The computation of \textcolor{blue}{$U_{(i,j),(s, j_s)}$} is done as follows.
\begin{enumerate}
\item If $\tau_{i}$ is a sink and subscription node. It means $\tau_{i}$ has only one immediate preceding producer, i.e., $producer(\tau_i)=\tau_p$.
\begin{itemize}
    \item If $\tau_{i}$ is directly connected to a sensor \textcolor{blue}{$\tau_s$ ($\tau_p=\tau_s$)}, every $j$-th instance of $\tau_i$ is triggered by $j$-th instance of \textcolor{blue}{$\tau_s$}. Hence,
    \textcolor{blue}{$U_{(i,j),(s, j)}=1$}
    \item
    If $\tau_i$'s producer, $\tau_p$, is a fusion node, we have: 
    \textcolor{blue}{$$U_{(i,j),(s, j_s)}=U_{(p,j),(s, j_s)}, \quad \forall \tau_s\in V,\theta_s=\textit{``sen''},j_s\in [1, \textit{n-ins}(\tau_s,\Delta)]$$} 
    This is obvious because any sensor instance \sout{$\tau_{i',j'}$}\textcolor{blue}{$\tau_{s,j_s}$} used by $\tau_{p,j}$ is in turn used by $\tau_{i,j}$.    
\end{itemize}

\item If $\tau_{i}$ is a fusion node (either sink or intermediate node). Since $\tau_{i}$ may have multiple preceding producers through $\tau_{i}$'s predecessors, we need to consider the following cases for all \sout{$\tau_p \in  \{producer(\tau_{i'})| \forall \tau_{i'} \in pred_i\}$}\textcolor{blue}{$\tau_p \in  pred\_producers(\tau_i)$.}
\begin{itemize}
    \item If $\tau_{p}$ is a sensor \sout{$\tau_{i'}$ ($\tau_p = \tau_{i'}$)}\textcolor{blue}{$\tau_s$ ($\tau_p = \tau_s$)}, we can directly use \sout{$u_{(i,j),(i',j')}$}\textcolor{blue}{$u_{(i,j),(s,j_s)}$}:
    \textcolor{blue}{$$U_{(i,j),(s, j_s)}=u_{(i,j),(s,j_s)},\quad \forall j_s\in [1, \textit{n-ins}(\tau_s,\Delta)]$$}
    \item If $\tau_{p}$ is a fusion node, we should take into account whether any instance of $\tau_{p}$ uses \sout{$\tau_{i',j'}$}\textcolor{blue}{$\tau_{s,j_s}$} or not. Also, we need to consider all paths that may exist between $\tau_i$ and \sout{$\tau_{i'}$}\textcolor{blue}{$\tau_s$}, as the DAG may contain branch nodes, resulting in multiple paths. These can be determined by the following lemma.
\end{itemize}
\end{enumerate}
\begin{lemma}[Constraint: $U$ for \underline{fusion} nodes] \label{lem:bigU_multiply}
The binary variable \textcolor{blue}{$U_{(i,j),(s,j_s)}$} for any fusion node $\tau_i$ and any sensor node \textcolor{blue}{$\tau_s$} is equal to 1 if the following inequality holds: \textcolor{blue}{
\begin{equation}\label{eq:bigU_multiply}
\sum_{\tau_p \in pred\_producers(\tau_i)}\sum_{j_p=1}^{\textit{n-ins}(\tau_p,\Delta)} u_{(i,j),(p,j_p)} \cdot U_{(p,j_p),(s, j_s)} \geq 1
\end{equation}}
where $j \in [1, \textit{n-ins}(\tau_i, \Delta)]$, \textcolor{blue}{and $j_s \in [1, \textit{n-ins}(\tau_s, \Delta)]$.}
\begin{proof}
Consider an instance $\tau_{i,j}$ that uses data from sensor instance \textcolor{blue}{$\tau_{s,j_s}$}. This data flows through an intermediate fusion node $\tau_p$.
For $\tau_{i,j}$ to use data from \textcolor{blue}{$\tau_{s,j_s}$}, there must exist some instance \textcolor{blue}{$j_p$} of $\tau_p$ such that: (i) $\tau_{i,j}$ uses \textcolor{blue}{$\tau_{p,j_p}$} (represented by \textcolor{blue}{$u_{(i,j),(p,j_p)}=1$}), and (ii) \textcolor{blue}{$\tau_{p,j_p}$} uses \textcolor{blue}{$\tau_{s,j_s}$ ($U_{(p,j_p),(s,j_s)}=1$)}. The product \textcolor{blue}{$u_{(i,j),(p,j_p)} \cdot U_{(p,j_p),(s, j_s)}$} equals 1 if and only if both conditions are met for instance \textcolor{blue}{$j_p$}.
The summation over all possible instances \textcolor{blue}{$j_p$} of $\tau_p$ captures every potential instance through which data from \textcolor{blue}{$\tau_{s,j_s}$} could reach $\tau_{i,j}$. Similarly, the summation over all $\tau_p \in pred\_producers(\tau_i)$ ensures that all possible paths in the DAG from \textcolor{blue}{$\tau_{s,j_s}$} to $\tau_{i,j}$ are considered. Therefore, \textcolor{blue}{$U_{(i,j),(s, j_s)} = 1$} if there exists at least one instance \textcolor{blue}{$j_p$} of $\tau_p$ on at least one path from \textcolor{blue}{$\tau_{s,j_s}$} to $\tau_{i,j}$, which is exactly what the $\geq 1$ condition represents.\qed
\end{proof}
\end{lemma}


With the calculation of $U$\sout{ now defined}, we can determine, for each sensor that has at least a path to a sink, exactly which instances of the sensor \textcolor{blue}{$\tau_s$} contribute to each instance of the sink node $\tau_\otimes$. However, multiple sensors may converge through fusion nodes to contribute data to a sink node. Among the end-to-end latency metrics discussed, MRT needs to identify the oldest data among all contributing sensor instances for a given sink instance $\tau_{\otimes,j}$, i.e., the earliest released sensor instance. To find this earliest release time among sensor instances contributing to a task instance $\tau_{i,j}$, we define the function $OTS(\tau_{i,j})$ as follows:
\begin{equation}\label{eq:ots}\textcolor{blue}{ \small
\begin{split}
OTS(\tau_{i,j}) &= min \big{\{}T_s \cdot (j_s-1) | \forall \tau_s \in V, \theta_s = \text{``sen''},\\ & \forall j_s \in [1, \textit{n-ins}(\tau_s, \Delta)], \quad \text{if } U_{(i,j),(s,j_s)}=1 \big{\}}   
\end{split}
}
\end{equation}
Similarly, to compute the MTD metric for a sink node, in addition to $OTS(\tau_{i,j})$, we must identify the newest data among all contributing sensor instances for the given sink instance $\tau_{\otimes,j}$. To capture this latest release time among sensor instances contributing to $\tau_{i,j}$, we define the function $NTS(\tau_{i,j})$ as follows:
\begin{equation}\label{eq:nts}\textcolor{blue}{ \small
\begin{split}
NTS(\tau_{i,j}) &= max \big{\{}T_s \cdot (j_s-1) | \forall \tau_s \in V, \theta_s = \text{``sen''},\\ & \forall j_s \in [1, \textit{n-ins}(\tau_s, \Delta)], \quad \text{if } U_{(i,j),(s,j_s)}=1 \big{\}}   
\end{split}
}
\end{equation}

Once we determine the $OTS$ and $NTS$ values for each sink instance, we can derive the reaction time and time disparity. To obtain the maximum values across all sink node instances within the interval $\Delta$, $OTS$ and $NTS$ need to be calculated for each instance. We define MRT following the first-to-first concept from \cite{durr2019end}, and MTD according to the worst-case time disparity definition in~\cite{jiang2023analysis}. Using our ILP model, these metrics for a sink node $\tau_{\otimes}$ are formulated as follows:
\begin{equation}\label{eq:mrt}
\text{\scalebox{0.9}{$MRT(\tau_\otimes) = \max\{f_{\otimes,j} - OTS(\tau_{\otimes,j-1}) | \forall j \in [1, \textit{n-ins}(\tau_{\otimes}, \Delta)]\}$}}    
\end{equation}
\begin{equation}\label{eq:mtd}
\text{\scalebox{0.9}{$MTD(\tau_\otimes) = \max\{NTS(\tau_{\otimes,j}) - OTS(\tau_{\otimes,j}) | \forall j \in [1, \textit{n-ins}(\tau_{\otimes}, \Delta)]\}$}}    
\end{equation}

To compute PAoI, we \sout{measure} \textcolor{blue}{determine} the time interval between the initiation of each sensor instance and its most recent update for every sensor contributing to a sink node. We then take the maximum interval across sensors per sink instance, and finally, the maximum among all sink instances. We formalize PAoI as follows\textcolor{blue}{, where $s_{s,j_s}$ denotes the start time of the sensor instance $\tau_{s,j_s}$}: 
\begin{equation}\textcolor{blue}{\small
\begin{split}
&PAoI(\tau_\otimes) = max\{(s_{s,j_s}-s_{s,j_s-1})\cdot U_{(\otimes,j),(s,j_s)}| \\
&\forall j_s \in [1, \textit{n-ins}(\tau_s, \Delta)], \quad \forall \tau_s \in V, \quad \forall j \in [1, \textit{n-ins}(\tau_\otimes, \Delta)] \}      
\end{split}
}
\end{equation}

The metrics above are formulated for each sink node for the entire DAG. In addition, we can determine the worst-case response time (WCRT) of individual chains from each sensor \sout{$\tau_{i'}$}\textcolor{blue}{$\tau_s$} to each sink node $\tau_\otimes$ as follows:
\begin{equation}\label{eq:wcrt}\textcolor{blue}{\small
\begin{split}
&WCRT(\tau_s, \tau_\otimes) = \max\{(f_{\otimes,j} - T_s \cdot (j_s-1))\cdot U_{(\otimes,j)(s, j_s)} | \\
&\forall j \in [1, \textit{n-ins}(\tau_{\otimes}, \Delta)], \quad \forall j_s \in [1, \textit{n-ins}(\tau_s, \Delta)]\}    
\end{split}
}
\end{equation}
Note that, unlike MRT, MTD, and PAoI, which consider all sensors contributing to an actuator, WCRT is specified for a single sensor. 

Another metric modeled in our work is the makespan (MS), often used to enhance resource efficiency. Makespan is the total time taken to complete a set of tasks, and in our model, it corresponds to the maximum finish time of a sink instance across all its instances. In ADS, however, where tasks can be triggered by events or on a timer, minimizing MS may seem less relevant since tasks cannot begin before their designated trigger times. Still, including MS as an optimization objective may provide insights for fine-tuning and calibrating the system. MS is formulated as follows for a sink node $\tau_{\otimes}$:
\begin{equation}\label{eq:mms} \small
MS(\tau_\otimes) = \max\{f_{\otimes,j}| \forall j \in [1, \textit{n-ins}(\tau_{\otimes}, \Delta)]\}  
\end{equation}

\noindent\textbf{\textcolor{blue}{Determining Time Interval $\Delta$.}}\label{sec:delta} \textcolor{blue}{Recall that we define the time interval $\Delta$ as $\Delta=k\cdot HP$, where $HP$ is the hyperperiod (Sec.~\ref{sec:model_fusion}). In conventional static schedule generation, $k=1$ is sufficient to check task-level schedulability because the same schedule will repeat afterwards. However, for end-to-end metrics like MRT, the initial hyperperiod serves as a warm-up phase \cite{teper2024end}, where all tasks and paths within the DAG complete at least one execution cycle (e.g., ensuring data availability when T-fusion and I-fusion nodes are triggered subsequently).
Therefore, in our ILP-based framework, we use $\Delta=3\cdot HP$ consisting of three phases: (i) first HP: a warm-up phase for the second HP and is excluded from performance evaluation; (ii) second HP: the optimized schedule that will repeat afterwards; (iii) third HP: an exact copy of the second HP's schedule with the same relative start and finish times for all task instances. The third HP is essential for evaluating end-to-end latency metrics (MRT, MTD, and PAoI), as data flows may span across hyperperiod boundaries. Hence, we optimize these metrics for two consecutive hyperperiods (the second and third) to capture cross-boundary data flows. Once an optimized schedule is determined, it can be applied to runtime systems by enforcing the predetermined start and finish times of the task instances and repeating the second hyperperiod's schedule.}

\smallskip
\sout{With these metrics formulated by our ILP model, we can now define objective functions that can be configured to \textit{minimize} each metric individually or as part of a multi-objective function, with configurable weights and priorities for Pareto optimality.}\textcolor{blue}{With these metrics formulated by our ILP model, we define our ILP objective function to \textit{minimize} these metrics using a hierarchical-blended approach \cite{hierarchical-blended}, where metrics are first optimized using lexicographic ordering by priority, and within the same priority level, a weighted sum is applied. This approach provides configurable priorities and weights that let us choose which metric(s) to focus on.
It is worth mentioning that while each metric is defined as a maximum (upper bound) value among different instances, our optimization objective is to minimize these maximum values. }


\section{Evaluation}\label{EVAL}
Our evaluation begins with case studies derived from existing literature, with extensions to capture additional fusion task types. We further evaluate our framework using randomly generated DAGs, varying the number of nodes, edges, and task types. The framework is implemented using Gurobi, a state-of-the-art ILP solver capable of efficiently handling linear, mixed-integer, and quadratic problems, and the optimal schedule from the framework has been tested on a Raspberry Pi. \sout{Throughout our evaluation, we use a time interval of $\Delta = 3 \cdot \text{hyperperiod}$ to find an optimal schedule (the rationale explained in Sec.~\ref{sec:delta} of the supplemental material; same schedule will repeat afterwards).} We also provide additional customized case studies in Sec.~\ref{sec:custom} of the supplemental material, focusing on unique features of our framework to highlight how our model handles diverse DAG configurations. \textcolor{blue}{Throughout this section, the relative deadlines of timer-triggered tasks are set equal to their periods, while for event-triggered tasks, the relative deadline is set to the largest period among the timer-triggered tasks. Additionally, the multi-objective function in our framework includes all metrics—MRT, MTD, PAoI, WCRT, and MS—with weights and priorities set to 1, unless stated otherwise.}

\subsection{Case Studies from \cite{teper2022end} and Beyond}
We compare our ILP-based framework to the approach proposed in \cite{teper2022end}, which focuses on optimizing MRT for chains within DAGs while supporting only certain types of fusion nodes. Their method represents one of the most advanced existing models for cause-effect chains and is evaluated using two case studies. To ensure a fair comparison, we align our setup by configuring the number of CPU cores to $\Pi = 1$, matching their single-core assumption. We apply our work to the same case studies and expand the task configurations beyond their model's support.

\begin{figure}[t]
\centering
\begin{minipage}[b]{0.4\linewidth}
    \centering
    \includegraphics[width=\linewidth]{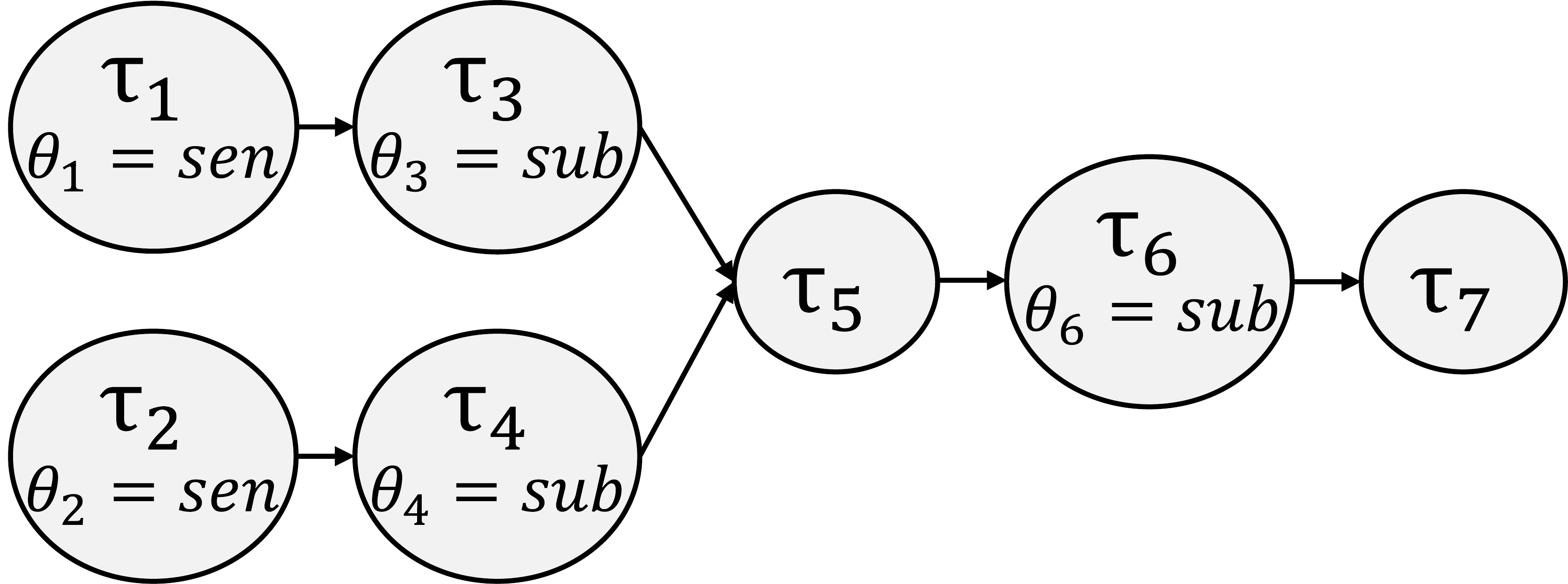}
    \subcaption{DAG structure}
    \label{fig:case1_general}
\end{minipage}
\hspace{0.2cm}
\begin{minipage}[b]{0.55\linewidth}
    \centering
    \subcaption{Task configurations}
    \resizebox{\linewidth}{!}{%
    \begin{tabular}{|c|c|c|c|} \hline
    Config & $\theta_5$ & $\theta_7$ & Task periods (ms) \\ \hline
    WS & ``w-fus'' & ``sub'' & $T_1 = T_2 = 360$\\ \hline
    WT & ``w-fus'' & ``t-fus'' & $T_1 = T_2 = 420$, $T_7=840$\\ \hline
    TS & ``t-fus'' & ``sub'' & $T_1 = T_2 = 420$, $T_5=840$\\ \hline
    TT & ``t-fus'' & ``t-fus'' & $T_1 = T_2 = 480$, $T_5=T_7=960$\\ \hline
    NH & ``t-fus'' & ``t-fus'' & $T_1 = 480$, $T_2=360$, $T_5=T_7=960$\\ \hline
    \end{tabular}
    }
    \label{tab:case1_configs}
\end{minipage}
\vspace{-3mm}
\caption{Fusion system with two chains}
\label{fig:case1}
\vspace{-5mm}
\end{figure}

\smallskip\noindent \textbf{Fusion System with Two Chains.} To facilitate comparison, we adopted the DAG structure shown in Fig.~\ref{fig:case1_general} from \cite{teper2022end} with minor adjustments. In \cite{teper2022end}, the authors explored two types of fusion nodes and two types of actuator nodes across four different configurations. In our setup, we mapped their ``subscription-fusion'' node type to our W-fusion node type; ``timer-fusion'' node to our T-fusion node; ``timer actuator'' node to our T-fusion sink node with a single predecessor; and ``subscription-actuator'' node to a subscription node in our model. In addition, to evaluate the MTD metric, we introduced a new configuration in which sensors have non-harmonic periods. Therefore, we used five configurations in total: \textit{(i)} WS: \underline{W}-fusion node with \underline{S}ubscription actuator, \textit{(ii)} WT: \underline{W}-fusion node with \underline{T}-fusion actuator, \textit{(iii)} TS: \underline{T}-fusion node with \underline{S}ubscription actuator, \textit{(iv)} TT: \underline{T}-fusion node with \underline{T}-fusion actuator, and \textit{(v)} NH: \underline{N}on-\underline{H}armonic periods for the TT configuration. The period $T_i$ and type $\theta_i$ of tasks that vary across configurations are listed in Table~\ref{tab:case1_configs}. For all configurations, we set the WCET of the tasks as follows: $e_1 = e_3 =10 (ms)$, $e_2 = e_4 =20 (ms)$, $e_5 = e_6 = e_7 = 30 (ms)$.

Table~\ref{tab:cs1_four_configs} summarizes the comparison results, where IDS indicates the results from our framework (\underline{I}LP-based \underline{D}AG \underline{S}chedule) when optimizing \sout{for a Pareto-optimal point across MRT, MTD, PAoI, and WCRT metrics}\textcolor{blue}{a multi-objective function of MRT, MTD, PAoI, and WCRT metrics}.
We include the analytical upper-bound (UB) and the simulation lower-bound (LB) on MRT computed using \cite{teper2022end}'s implementation\footnote{\cite{teper2022end} analyzes only the MRT of each chain. Hence, we report the maximum MRT upper-bound and lower-bound across all chains in the DAG.}; however, note that these are for reference only as \cite{teper2022end} analyzes MRT bounds under ROS2 scheduling while our framework finds the performance bounds of various metrics through an optimal schedule. The results show that our optimal schedule can achieve substantially lower MRT compared to conventional scheduling while also optimizing other metrics. MTD is not reported in the table since it is zero for the first four configurations due to the same sensor periods, and equals 120 ms for the NH configuration.

\begin{table}[t]
\centering
\caption{\textcolor{blue}{MRT comparison with prior work and PAoI/WCRT results}
} 
\scriptsize
\begin{tabular}{|c|c|c|c|c|c|}
\hline
\textbf{Config} & \textbf{\cite{teper2022end}'s UB} & \textbf{\cite{teper2022end}'s LB} & \textbf{IDS:MRT} & \textbf{IDS:PAoI} & \textbf{IDS:WCRT of chain 1} \\ \hline
WS & 2190 & 510 & 510 & 360 & 150 \\ \hline
WT & 3780 & 1320 & 990 & 120 & 150 \\ \hline
TS & 2420 & 1410 & 990 & 60 & 150 \\ \hline
TT & 3830 & 2490 & 1110 & 60 & 150 \\ \hline
NH & 3830 & 2490 & 1250 & 40 & 250 \\ \hline
\end{tabular}
\label{tab:cs1_four_configs}
\end{table}

\begin{figure}[t]
\centering
\scalebox{0.8}{
\begin{minipage}[b]{0.30\linewidth}
    \centering
    \includegraphics[width=\linewidth]{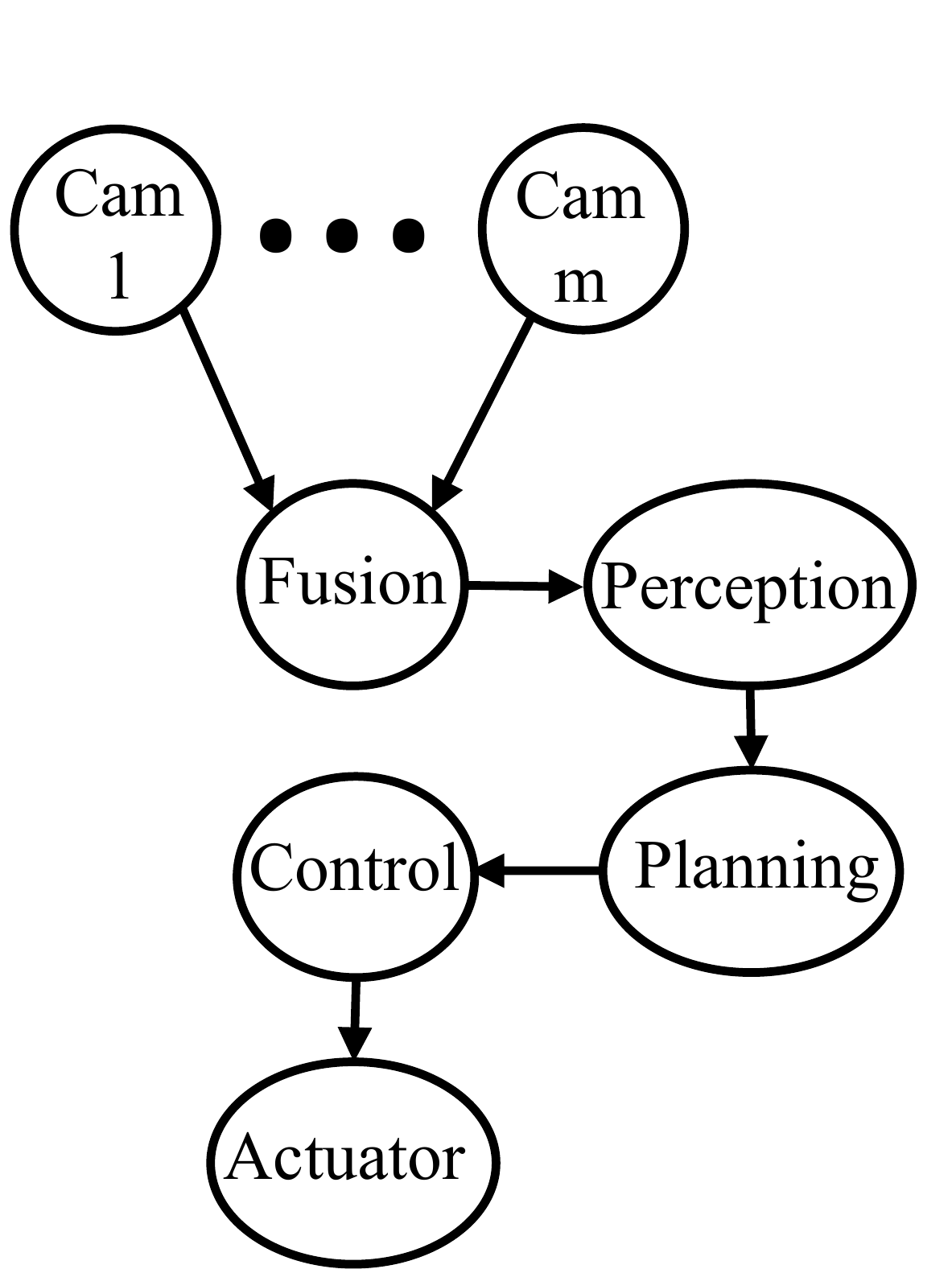}
    \subcaption{Navigation system with $m$ cameras}
    \label{fig:navi_dag}
\end{minipage}
\hspace{1cm}
\begin{minipage}[b]{0.65\linewidth}
    \centering
    \subcaption{{MRT comparison between \cite{teper2022end} (UB and LB) and our IDS output executed on Raspberry Pi 4}}
    \resizebox{\linewidth}{!}{%
    \begin{tabular}{|c|c|c|c|c|}
    \hline
    $m$ & \cite{teper2022end}'s UB & \cite{teper2022end}'s LB & IDS:MRT & observed MRT (ms) \\ \hline
    1 & 480 & 155 & 155 & 155.06 \\ \hline
    2 & 515 & 160 & 160 & 159.85 \\ \hline
    3 & 840 & 165 & 165 & 164.94 \\ \hline
    4 & 890 & 170 & 170 & 170.37 \\ \hline
    5 & 940 & 175 & 175 & 175.38\\ \hline
    6 & 990 & 180 & 180 & 180.31 \\ \hline
    7 & 1040 & 185 & 185 & 184.94 \\ \hline
    8 & 1090 & 190 & 190 & 190.48 \\ \hline
    9 & 1140 & 195 & 195 & 194.93 \\ \hline
    10 & 1190 & 530 & 200 & 200.01 \\ \hline
    \end{tabular}    
    }
    \label{tab:navi_teper_vs_rpi}
\end{minipage}
}
\caption{Navigation System Case Study}
\label{fig:navi}
\end{figure}

\smallskip\noindent \textbf{Navigation System.} This case study from \cite{teper2022end}, inspired by a navigation system, models its DAG as shown in Fig.~\ref{fig:navi_dag}. The number of cameras can vary, allowing us to assess the impact of scaling on system performance. \textcolor{blue}{We denote the optimal result from our framework as IDS.} We \textcolor{blue}{also} report the MRT value obtained by implementing our optimal schedule as a static user-level scheduler in Linux and running it for 100 hyperperiods on a Raspberry Pi 4 (64-bit quad-core ARM Cortex-A72)\textcolor{blue}{, denoted as the observed MRT}. We then compare this against the maximum UB and LB on MRT across all chains computed by \cite{teper2022end}, as the number of cameras increases. To align with their model, we map their fusion node to our W-fusion type. We adopt the same WCET values as in \cite{teper2022end} for all other nodes. Table~\ref{tab:navi_teper_vs_rpi} shows that our observed MRT matches their simulation LB at lower camera counts and stays lower when their MRT increases at 10 cameras. \textcolor{blue}{Results for other metrics are as follows: MTD = 0 and PAoI = 100 ms for all camera counts (due to fixed camera periods); MS increases from 255 ms to 300 ms; and WCRT increases from 55 ms to 100 ms, both rising in 5 ms increments per additional camera.} We introduce additional configurations beyond \cite{teper2022end}'s case study, where the fusion node is a T-fusion or I-fusion. Due to space limits, the results are provided in the supplemental material (Sec.~\ref{sec:navi_contd}). 


\subsection{Randomly-Generated DAGs}
To evaluate our framework with more complex DAGs with diverse fusion types, we use randomly generated DAGs. 
We conducted this set of experiments on a Linux server equipped with two AMD EPYC 7452 processors (32 cores each), providing a total of 64 cores and 256GB (16GB x 16) of DDR4 RAM.

The size of each experiment is set to 100 randomly generated DAGs. The WCET of event-triggered tasks was randomly selected from $[1, 5] (ms)$. For timer-triggered tasks, we assigned a random utilization from $[0.1, 0.4]$ and chose a period from $\{20, 40, 50, 100\} (ms)$, calculating their WCET as the product of utilization and period. The number of cores used for scheduling is set to $\Pi=2$. \sout{We configured our ILP method to optimize multiple objectives—MRT, PAoI, and WCRT—assigning equal weight and priority to each.} Since our model supports multiple sink nodes, we calculated MRT\textcolor{blue}{, MTD,} and PAoI for the sink with the last index. For WCRT, defined along a chain from a sensor to a sink node, we focused on the path between the sensor with the first index and the sink with the last index.

\begin{figure}[t]
\centering
\begin{minipage}[b]{0.41\linewidth}
    \centering
    \includegraphics[width=\linewidth, height=2.6cm]{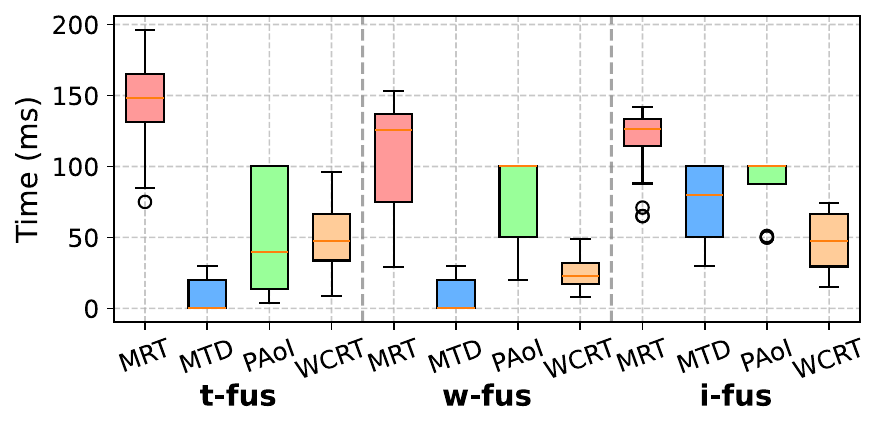}\vspace{-2mm}
    \subcaption{Varying the allowed fusion types}
    \label{fig:varying_fus}
\end{minipage}
\hfill
\begin{minipage}[b]{0.55\linewidth}
    \centering
    \includegraphics[width=\linewidth]{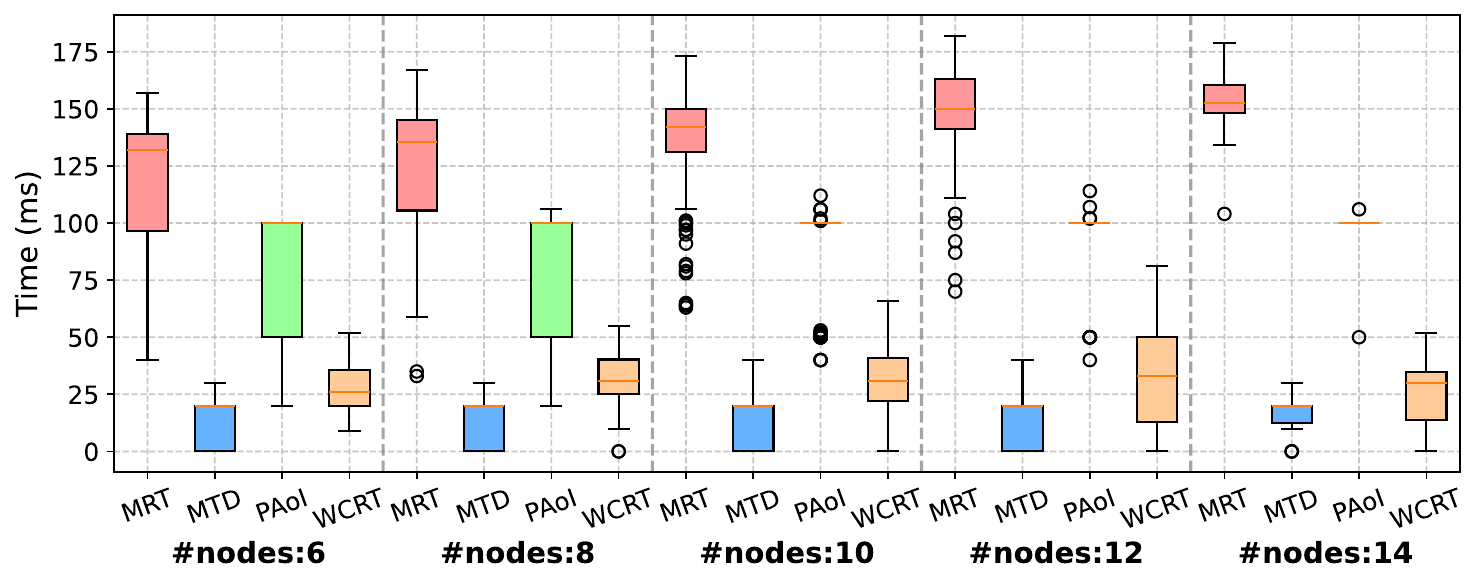}\vspace{-2mm} 
    \subcaption{\scriptsize{Varying \#nodes and \#edges (W-fusion allowed)}}    
    \label{fig:varying_nodes}
\end{minipage}
\begin{minipage}[b]{0.96\linewidth}
    \centering
    \includegraphics[width=0.9\linewidth]{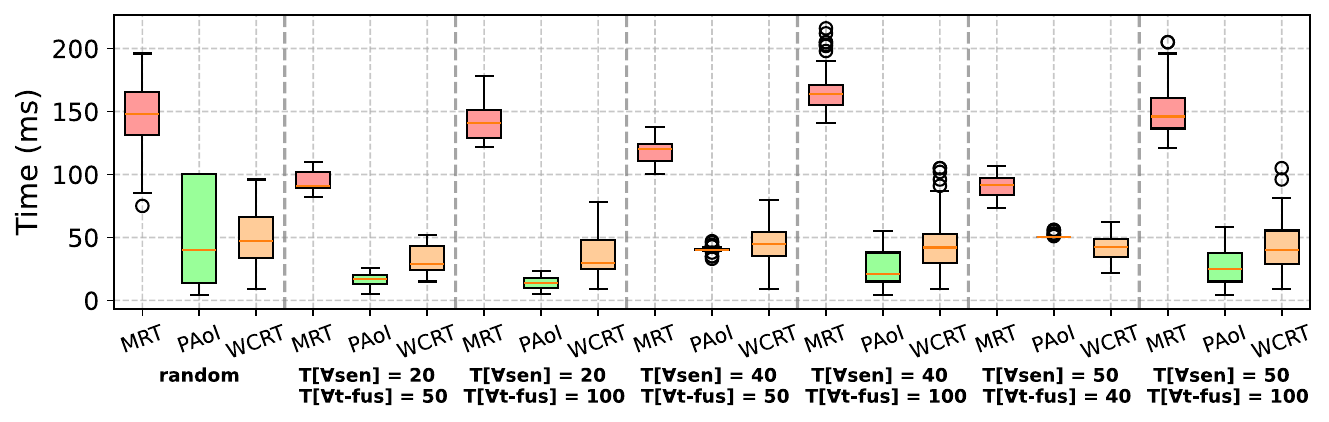}\vspace{-2mm}
    \subcaption{Varying the periods (T-fusion allowed)}
    \label{fig:varying_periods}
\end{minipage}
\vspace{-3mm}
\caption{{\small Distribution of optimized real-time performance metrics across feasible cases}}
\label{fig:distribution}
\end{figure}

We first explored the impact of fusion types on a set of 100 DAGs, each consisting of 6 nodes — 3 of which are sensor nodes — and 7 edges. Using these same DAG structures, we varied the fusion node type by allowing multiple fusion nodes, but all of the same type — either T-fusion, W-fusion, or I-fusion — within each DAG. 
The distribution of optimized metrics across the feasible cases (out of 100) is shown in Fig.~\ref{fig:varying_fus}, highlighting the expected effects of each fusion type. \sout{For example, comparing MRT across fusion types shows that T-fusion can result in high MRT if periods are not carefully chosen, while I-fusion tends to reduce reaction time by triggering on each input.}
\textcolor{blue}{For example, comparing MRT and MTD across fusion types shows that I-fusion tends to reduce MRT by acting on each new input immediately. However, this same behavior increases MTD as it leads to multiple reuses of some inputs, increasing the deviation between the oldest and newest sensor data used for an actuator. We can also observe that T-fusion may result in high MRT if task periods are not carefully chosen. To better investigate how task periods affect these metrics, we conducted another experiment in which only T-fusion is allowed for fusion nodes, assuming task periods can be configured at design time, as shown in Fig.~\ref{fig:varying_periods}. We compared the default case where sensor and T-fusion node periods are randomly selected from \{20, 40, 50, 100\}(ms) to cases with fixed period assignments. For simplicity, we assume all sensors (and T-fusion nodes) share the same period, denoted as $T[\forall sen]$ (and $T[\forall \textit{t-fus}]$) in the figure. We excluded MTD from the results since it is zero in all cases except the random one, due to uniform sensor and T-fusion periods. The results show that there exist cases where the random case has better MRT than those with T-fusion periods much shorter than sensor periods. In fact, MRT improves when sensor and T-fusion periods are closely aligned. We also found that lower PAoI occurs when sensors have shorter periods and sample data more frequently. In this experiment, we can see that if periods can be tuned, then our framework gives valuable insight on their effect on real-time metrics.
}

\begin{figure}[t]
    \centering
    \subfloat[\textcolor{blue}{Schedulability ratio}]{\includegraphics[width=0.4\linewidth]{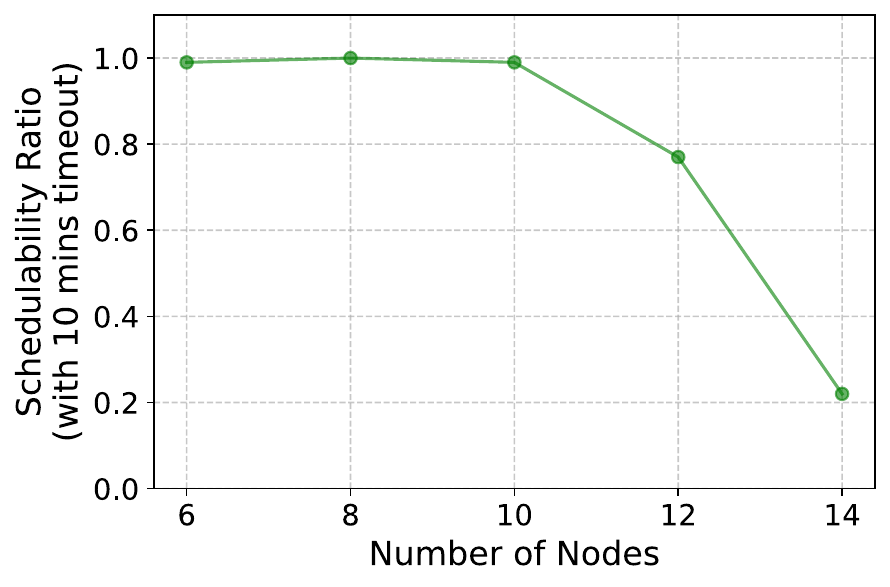}\label{fig:success_rate}\vspace{-2mm}}
    \hspace{0.05\linewidth} 
    \subfloat[Average runtime]{\includegraphics[width=0.4\linewidth]{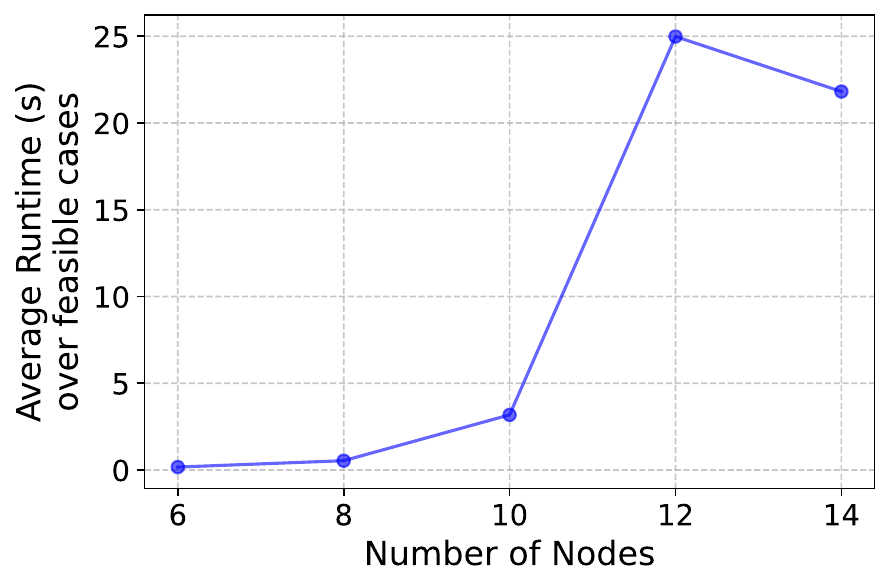}\label{fig:avg_runtime}\vspace{-2mm}}
    \vspace{-2mm}
    \caption{The \textcolor{blue}{schedulability ratio (10-minute timeout)} and average runtime \textcolor{blue}{per case} of our framework}
    \label{fig:success_runtime}
\end{figure}

In another experiment, we evaluated our framework by increasing DAG complexity through varying the number of nodes and edges. We considered \sout{six}\textcolor{blue}{five} configurations, assuming the number of sensor nodes to be half the total nodes and the number of edges to be twice the nodes. The number of edges was kept within the valid DAG range, $[|V|- 1, |V|\cdot (|V|-1)/2]$ where $|V| = m + n$ is the total number of nodes. For each configuration, we generated 100 DAGs and analyzed the distribution of real-time metrics, as shown in Fig.~\ref{fig:varying_nodes}. In this experiment, we fixed the fusion type to W-fusion, except for I-fusion following branch nodes. \sout{The results confirm our previous observations and further}\textcolor{blue}{The results} show that real-time metrics, such as MRT, increase with DAG complexity, while the number of feasible cases declines due to the growing workload on a fixed number of cores ($\Pi=2$). We further evaluated the \sout{success rate}\textcolor{blue}{schedulability ratio} and the average runtime of our framework using the same configurations. \sout{The success rate is defined as the ratio of feasible cases returned by our framework out of 100 DAGs, with a 10-minute timeout for infeasible cases.} \textcolor{blue}{The schedulability ratio is the ratio of feasible (schedulable) cases returned by our framework within 10 minutes out of 100 DAGs.} 
As shown in Fig.~\ref{fig:success_runtime}, both the \sout{success rate} \textcolor{blue}{schedulability ratio} and the average runtime for feasible cases are influenced by the increasing complexity of the DAGs, while the number of resources remains unchanged. \sout{The smaller runtime in Fig.~\ref{fig:avg_runtime} and the lower timing metrics in Fig.~\ref{fig:varying_nodes} for 16 nodes compared to 14 nodes are both due to the decreased number of feasible cases.}\textcolor{blue}{Note that the smaller runtime observed for 14 nodes compared to 12 nodes in Fig.~\ref{fig:avg_runtime}, as well as the lower MTD and PAoI values for 14 nodes than those for 12 nodes in Fig.~\ref{fig:varying_nodes}, are both due to the fewer feasible cases.}

To  further assess the flexibility of our framework, we increased the number of cores, which enabled scheduling even more complex DAGs with improved success rates and scalable average runtimes. The results of this extended evaluation are provided in the supplemental document (Sec.~\ref{sec:eval-cores}) due to space limits.


\section{Conclusion}
In this paper, we introduce flexible and structured modeling for data fusion tasks in the Autonomous Driving System (ADS) software stack, supporting complex task types and chains with diverse triggering options \sout{not addressed by existing models}\textcolor{blue}{that existing models have not addressed comprehensively and systematically}. We present an ILP-based framework that quantitatively compares different fusion patterns and their impact on real-time performance metrics, providing optimized resource allocation (task-to-core mapping) and timing schemes for all task instances. Our framework can optimize various real-time performance metrics, such as Maximum Reaction Time (MRT), Maximum Time Disparity (MTD), Peak Age of Information (PAoI), Worst-Case Response Time (WCRT), and Makespan (MS), allowing users to adjust fusion strategies and other system parameters. Evaluation against existing approaches demonstrates that our framework not only better handles complex DAG structures found in real systems, but also effectively analyzes achievable bounds for key performance metrics. \textcolor{blue}{In the future, we plan to improve the scalability of our framework by incorporating heuristics such as simulated annealing and learning-based techniques for highly complex ADS.} 
\section*{Acknowledgment}
This work was sponsored by the National Science Foundation (NSF) grants 1943265 and 2312395.



\bibliographystyle{splncs04}
\bibliography{references}

\input{appendix}


\end{document}

%% file: appendix.tex
\thispagestyle{empty}  
\null  
\clearpage  

\appendix
\setcounter{page}{1}

{\centering \Large \textbf{Supplemental Material: Modeling and \sout{Timing Analysis}\textcolor{blue}{Scheduling} of Fusion Patterns in Autonomous Driving Systems}}

\section{ILP Formulation of General Scheduling Requirements} \label{sec:general}

This document explains how to formulate the general requirements of multi-core real-time task scheduling in ILP constraints. \sout{To determine task instances' start and finish times, we present two alternative formulations: (i) one that checks scheduling decisions at each discrete time point, and (ii) another that directly prevents overlapping executions between any pair of task instances.
For simplicity, we use $\Delta$ as a general notation for a time interval of interest and will specify how it should be determined later.}\textcolor{blue}{We first present the formulation for assigning task instances to cores.
Next, we provide formulations to determine the start and finish times of task instances, ensuring that no two task instances overlap on the same core, as well as formulations to determine their release times. Later in this section, we elaborate on how our framework manages branch-then-fusion paths within a DAG.}

\subsection{Instance to Core Assignment}\label{sec:mapping}
The DAG under analysis, $G$, comprises $m$ sensor nodes and $n$ non-sensor nodes, with a total size of $|V| = m + n$. This DAG operates on a multi-core platform equipped with $\Pi$ identical CPU cores. 
\sout{Assuming that tasks can be globally scheduled across cores, meaning that instances of a task may execute on different cores,} 
\textcolor{blue}{As noted in Sec.~\ref{sec:formulation}, We do not restrict tasks to be bound to specific cores, allowing instances of the same task to execute on different cores. Hence, }
we define a binary variable $y^{\pi}_{i,j}$ to indicate whether an instance $\tau_{i,j}$ is assigned to core $\pi$ (where $1 \leq \pi \leq \Pi$).
\begin{equation}\label{eq:y_def}
y^{\pi}_{i,j}= \begin{cases}
1, & \text{if instance $\tau_{i,j}$ is assigned to core $\pi$} \\
0, & \text{otherwise}
\end{cases}
\end{equation}

\textcolor{blue}{Note that allowing task instances to run on different cores does not necessarily mean that the schedule generated by our work falls into the global scheduling category. While the distinction between global and partitioned scheduling is typically used to describe runtime scheduler behavior, our work is different: if the optimal solution is to assign all instances of the same task to the same core, our framework will generate such a schedule. Conversely, if someone wants to enforce that all instances of the same task run on the same core, they can simply add an additional constraint using $y_{i,j}^{\pi}$, i.e., for each task $\tau_i$, all $y_{i,j}^{\pi}$ values must be the same across all instances $j$ for the core $\pi$.}

Since an instance can only be assigned to one core for execution, the following lemma holds:

\begin{lemma}[Constraint: Instance assignment]\label{lem:sum_y_over_p}
To ensure that the instance $\tau_{i,j}$ is assigned to exactly one core $\pi$, the following constraint must be satisfied:
\begin{equation}
\sum_{\pi \in \{1, \dots, \Pi\}} y^{\pi}_{i,j} = 1, \quad \forall \tau_i \in V, \forall j \in [1, \textit{n-ins}(\tau_i, \Delta)]    
\end{equation}

\begin{proof}
Obvious by definition of $y^{\pi}_{i,j}$.
\end{proof}
\end{lemma}

In a scheduling problem, once a task instance is assigned to a core, its start and finish times need to be determined. We denote the start time and finish time for a task instance $\tau_{i,j}$ as $s_{i,j}$ and $f_{i,j}$, respectively. Since tasks are non-preemptible, each task runs uninterrupted once it starts. Therefore, given the start time, we can calculate the finish time by adding the WCET $e_i$ of the task as follows:
\begin{equation}\label{eq:f_s_e} f_{i,j} = s_{i,j} + e_i \end{equation}

This simplifies our scheduling problem: we only need to determine each task instance's start time, not both start and finish times. \textcolor{blue}{Once the start time is determined, the finish time follows directly, allowing us to verify that each instance finishes before its deadline $d_{i,j}$, i.e., $f_{i,j} \leq d_{i,j}$. The absolute deadline of an instance $\tau_{i,j}$ is obtained by adding the task’s relative deadline $D_i$ to the time at which the instance is released and ready to start, denoted as $r_{i,j}$, giving $d_{i,j} = r_{i,j} + D_i$. In the following sections, we first explain how to determine start times, followed by the details of determining task release times.}

\subsection{Instance-Pair Based Formulation}\label{sec:continuous}  
\sout{While the time-point based formulation using $x_{i,j}^{\pi,t}$ works, its complexity can multiply with the length of $\Delta$. Hence, we present an alternative that directly compares any pair of two task instances. To do so, we use a different binary variable, $z_{(i,j),(i',j')}$, which indicates the execution order between instances of two different tasks.}
\textcolor{blue}{In this section, to determine the start time of task instances, we compare any pair of task instances to ensure there is no overlap in their execution on the same core. To achieve this, we use a binary variable $z_{(i,j),(i',j')}$, which indicates the execution order between instances of two different tasks.}
\begin{equation}\label{eq:z_def}
z_{(i,j),(i',j')}= \begin{cases}
1, & \text{\small{if instance $\tau_{i,j}$ executes before $\tau_{i',j'}$}} \\
0, & \text{otherwise}
\end{cases}
\end{equation}

Based on this definition, we present the following lemma\sout{ as an alternative to Lemma \ref{lem:no_overlap} for the continuous-time model}:

\begin{lemma}[Constraint: No overlapping executions on the same core]\label{lem:no_overlap_continuous_time}
For any pair of instances $\tau_{i,j}$ and $\tau_{i',j'}$, the following constraints must be satisfied to prevent overlapping executions on a core $\pi$: 
\begin{equation} \begin{split}
&\text{\scalebox{0.9}{
$s_{i',j'} \geq f_{i,j} - M(1-y^{\pi}_{i,j}) - M(1-y^{\pi}_{i',j'}) - M(1-z_{(i,j),(i',j')})$}}\\
&\text{\scalebox{0.9}{
$s_{i,j} \geq f_{i',j'} - M(1-y^{\pi}_{i,j}) - M(1-y^{\pi}_{i',j'}) - M\cdot z_{(i,j),(i',j')}$}}
\end{split}
\end{equation}
where $M$ is a big number ($M > $ hyper-period), and the hyper-period is defined as the least common multiple of the periods of all periodic tasks within the DAG.
\begin{proof}
These constraints use the Big-M method. If the instances are on different cores, either $y^{\pi}_{i,j}=0$ or $y^{\pi}_{i',j'}=0$, deactivating both constraints (e.g., $s_{i,j} \geq$ a negative value, trivially true). If both instances are on the same core ($y^{\pi}_{i,j} = y^{\pi}_{i',j'} = 1$), and $\tau_{i,j}$ executes before $\tau_{i',j'}$ ($z_{(i,j),(i',j')} = 1$), the first constraint becomes $s_{i',j'} \geq f_{i,j}$, and the second constraint becomes $s_{i,j} \geq f_{i',j'} - M$ (inactive). Conversely, if $\tau_{i',j'}$ executes first ($z_{(i,j),(i',j')} = 0$), the first constraint becomes $s_{i',j'} \geq f_{i,j} - M$ (inactive), and the second becomes $s_{i,j} \geq f_{i',j'}$. This ensures non-overlapping execution for instances on the same core.
\end{proof}
\end{lemma}

\subsection{\textcolor{blue}{Release Time Formulation}}\label{sec:release} 
\textcolor{blue}{In this section, we briefly discuss how the release time is determined for each task type.}

\textcolor{blue}{For timer-triggered tasks, such as sensor and T-fusion nodes, determining the release time is straightforward since they follow conventional periodic task behavior. The release time for a ``sen'' or ``t-fus'' task instance $\tau_{i,j}$ is given by: $$r_{i,j} = T_i \cdot (j-1)$$}

\textcolor{blue}{For event-triggered tasks, a task instance $\tau_{i,j}$ is released once certain instance(s) of its predecessor(s) have finished. For a subscription task, which has only one predecessor and inherits the instance index of that predecessor, the release time of the ``sub'' task instance $\tau_{i,j}$ is given by: $$r_{i,j} = f_{i', j} \quad where \quad pred_i =\{\tau_{i'}\}$$}

\textcolor{blue}{For W-fusion tasks, by definition, a new task instance $\tau_{i,j}$ is triggered and released once certain instances from all its predecessors have arrived. Since the selection of predecessor instances used by $\tau_{i,j}$ is explained in Sec.~\ref{sec:formulate_fusion}, we use the binary variable $u$ to define the release time of a ``w-fus'' task instance as follows: $$r_{i,j} = max \big{\{} f_{i',j'} \cdot u_{(i,j),(producer(\tau_{i'}), j')} | \forall \tau_{i'} \in pred_i, \forall j' \in [1, \textit{n-ins}(\tau_{i'}, \Delta)] \big{\}}$$}

\textcolor{blue}{For I-fusion tasks, except for the first instance, a new task instance $\tau_{i,j}$ is triggered and released once a certain instance of any one of its predecessors has finished. For the first instance of an ``i-fus'' task, i.e., $\tau_{i,1}$, the release time is defined as $r_{i,1} = max \big{\{} f_{i',1} | \forall \tau_{i'} \in pred_i \big{\}}$. For subsequent instances $\tau_{i,j}$ with $j \geq 2$, the release time is given by: $$r_{i,j} = f_{i',j'} \cdot u_{(i,j),(producer(\tau_{i'}), j')} \quad where \quad \tau_{i'} \in pred_i \land j' \in [1, \textit{n-ins}(\tau_{i'}, \Delta)] $$}

\subsection{\textcolor{blue}{Modeling Subscription Tasks in Branch-then-Fusion Structures}}\label{sec:branch} 


\textcolor{blue}{In our framework, a branch node and its successors can be any of the types described in Sec.~\ref{sec:model_fusion}. However, if any immediate successors of a branch node are subscription tasks, we treat these subscription tasks as a special case of an I-fusion node with a single predecessor for formulation purposes. This adjustment is necessary because, in our model, a subscription task inherits the instance index of its preceding sensor or fusion task ($producer$, defined in Sec. \ref{sec:formulate_fusion}) and does not generate new instance indices. This prevents different paths in a branch-then-fusion structure from having distinct instance indices, which is needed when some paths get input from a sensor or yet another fusion node. To be more precise, recall the binary variable $u_{(i,j)}{(p, j_p)}$ given in Def.~\ref{def:u} that links between a fusion task instance $\tau_{i,j}$ and the producer of one of its predecessors $\tau_{p, j_p}$ where $\tau_p \in pred\_producers(\tau_i)$. If $\tau_p$ branches out to multiple paths and contributes to $\tau_i$ through more than one path, the use of $u_{(i,j)}{(p, j_p)}$ can determine only the $j_p$-th instance of $\tau_p$ that reaches $\tau_{i,j}$ via one path, while other contributing instances of $\tau_p$ via other paths cannot be considered. This necessitates creating separate instance indices for the successors of the branch node. 
}

\textcolor{blue}{If we model the subscription tasks that are immediate successors of the branch node as I-fusion nodes, when finding $pred\_prodcers(\tau_i)$, each of these adjusted nodes will be included in $pred\_prodcers(\tau_i)$ as a $producer$. The branch node itself will only appear in $pred\_prodcers(\tau_i)$ if it is also an immediate predecessor of $\tau_i$. The reason for choosing I-fusion instead of T-fusion or W-fusion is that I-fusion generates a new instance index whenever the branch node produces a new output. Therefore, our framework correctly models the branch-then-fusion paths in a DAG.}

\section{Additional Evaluation}\label{sec:eval-contd}
In this section, we present additional experiments that further support our earlier observations and demonstrate how our framework handles different task types and configurations.

\subsection{Navigation System Case Study – Extended Configurations}\label{sec:navi_contd}
\begin{figure}[t]
\centering
\includegraphics[width=0.6\linewidth]{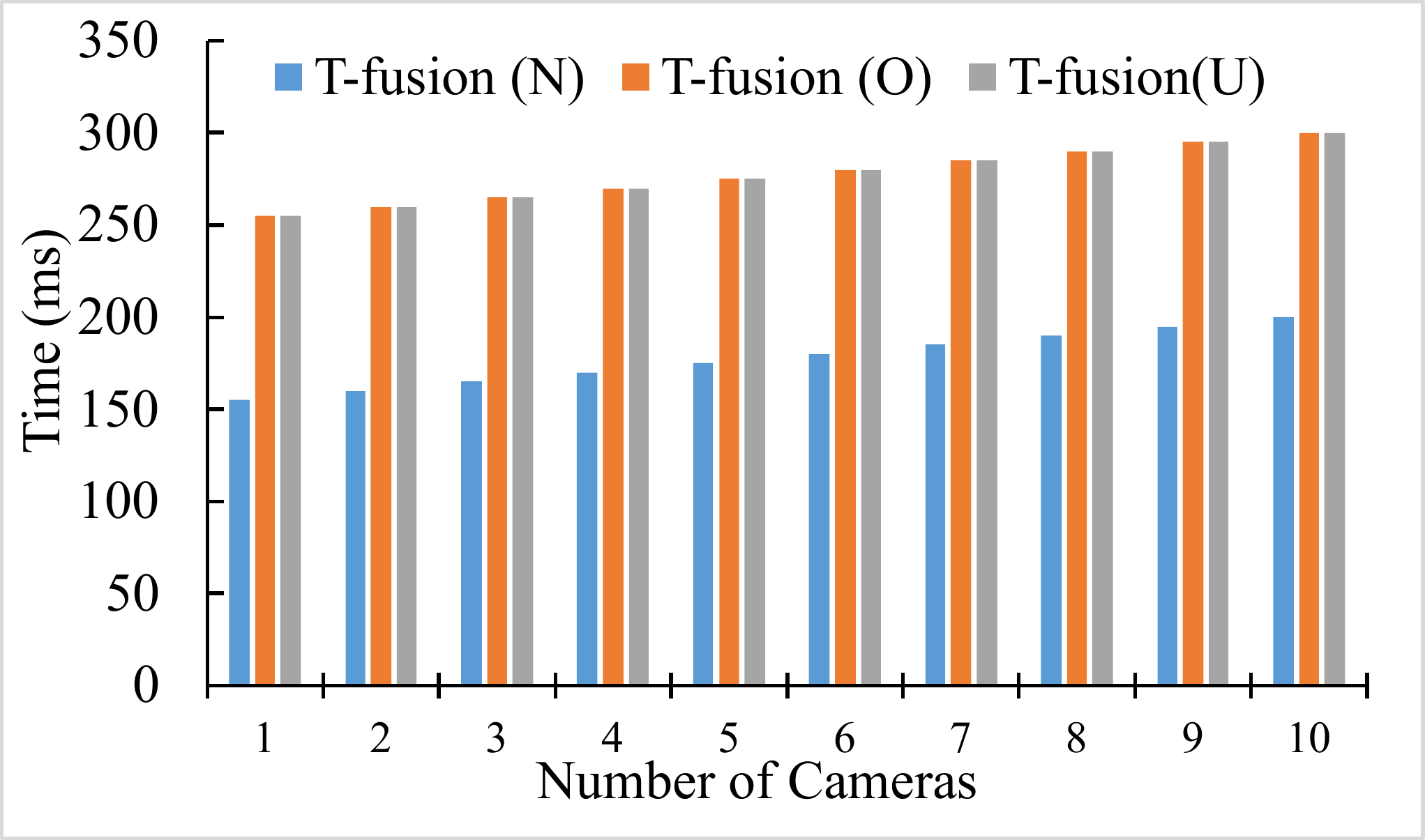}
\caption{MRT for three workloads with T-fusion node}
\label{fig:navi_tfus_mrt}
\end{figure}

For the navigation system case study shown in Fig.~\ref{fig:navi_dag}, we explore additional configurations beyond \cite{teper2022end}, where the fusion node is replaced with either a T-fusion or an I-fusion node. In the T-fusion setup, we examine how varying the T-fusion period influences system performance, offering insights into optimal system tuning.  We define three workload settings based on the periods of the cameras and T-fusion node: \textit{(a)} Normal (N): where all camera and T-fusion periods are equal, \textit{(b)} Over-sampled (O): where the T-fusion period is twice that of the cameras, and \textit{(c)} Under-sampled (U): where the T-fusion period is half that of the cameras. Fig.~\ref{fig:navi_tfus_mrt} present the MRT of the actuator across different workloads as the number of cameras increases. As anticipated, the MRT is higher in the under-sampled workload compared to the normal configuration. 

We also tried examining the same DAG by replacing the T-fusion node with an I-fusion node. However, doing so makes it feasible only when the system includes a single camera. This result is expected, as I-fusion nodes trigger each time a new camera instance arrives. Such frequent triggering causes the perception, planning, control, and actuator nodes to execute multiple times, consuming a significant portion of system resources. Consequently, the high cumulative WCET of these nodes restricts the number of I-fusion executions the system can accommodate on a single-core platform.

\subsection{Custom Case Studies on Model Features}\label{sec:custom}
In this section, we focus on case studies not modeled in prior work. We begin with the I-fusion node, which, as explained in the previous section, consumes significant resources due to being triggered by every sensor data arrival. Despite its complexity, this node type is commonly found in ADS and can improve MRT in some cases. Using a simple DAG with two sensors and one fusion node, we compare the MRT and MTD of an I-fusion sink node against a W-fusion sink node across two example task settings, demonstrating that I-fusion does not always lead to increased MRT (see Table.~\ref{tab:ifus_better_wfus}). In fact, in these examples, I-fusion achieves lower MRT. However, the MTD comparison reveals a trade-off: although I-fusion achieves lower MRT in these cases, it may rely on older data, resulting in a greater disparity between the timestamps of its inputs.

\begin{table}[t]
\centering
\caption{{\small Comparison of performance metrics with I-fusion and W-fusion nodes}}
\resizebox{0.8\columnwidth}{!}{%
\begin{tabular}{|c|c|c|c|c|} 
\hline
\multirow{2}{*}{\textbf{Settings}} & \textbf{$\theta_3$ = ``i-fus''} & \textbf{$\theta_3$ = ``i-fus''} & \textbf{$\theta_3$ = ``w-fus''} & \textbf{$\theta_3$ = ``w-fus''}  \\ 
\cline{2-5}
 & MRT & MTD & MRT & MTD \\ \hline
$\tau_1 = \{1, 5, \text{``sen''}, \varnothing\}$ & \multirow{4}{*}{9} & \multirow{4}{*}{6} & \multirow{4}{*}{12} & \multirow{4}{*}{2}  \\ 
$\tau_2 = \{1, 7, \text{``sen''}, \varnothing\}$ &                    &                    &                    &                     \\ 
$\tau_3 = \{1, 0, \theta_3, [\tau_1, \tau_2]\}$ &                    &                    &                    &                     \\  
$\Pi=1$ &                    &                    &                    &  \\                    
\hline
$\tau_1 = \{1, 3, \text{``sen''}, \varnothing\}$ & \multirow{4}{*}{6} & \multirow{4}{*}{3} & \multirow{4}{*}{8} & \multirow{4}{*}{1}  \\ 
$\tau_2 = \{1, 4, \text{``sen''}, \varnothing\}$ &                    &                    &                    &                     \\ 
$\tau_3 = \{1, 0, \theta_3, [\tau_1, \tau_2]\}$ &                    &                    &                    &                     \\  
$\Pi=2$ &                    &                    &                    &  \\ 
\hline
\end{tabular}
}
\label{tab:ifus_better_wfus}
\end{table}

\begin{figure}[t]
\centering
\subfloat[DAG with branch and fusion nodes]{\includegraphics[width=0.25\linewidth]{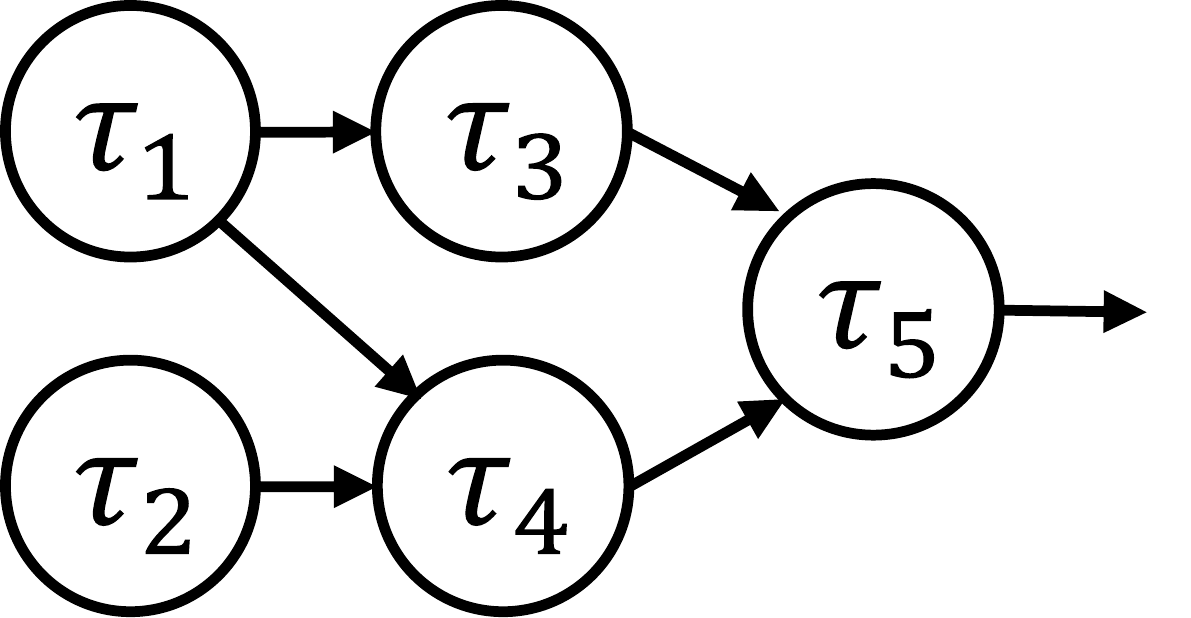}\label{fig:fork_fig}}\hspace{1cm}
\subfloat[Results of multi-objective optimization across MRT, MTD, and PAoI]{
\resizebox{0.45\linewidth}{!}{%
\begin{tabular}{|c|c|c|c|c|} 
\hline
Configurations & MRT & MTD & PAoI  \\ \hline
A & 37 & 0 & 20 \\
B & 60 & 5 & 17\\ \hline
\end{tabular}
}\label{fig:fork_results}}
\caption{Branch Node Case Study}
\label{fig:fork}
\end{figure}

We further examine a DAG containing both a branching node and a fusion node to test the robustness of our ILP approach when multiple paths link a sensor to an actuator. This experiment aims to validate the accuracy and flexibility of our framework in managing complex task dependencies. Using the DAG illustrated in Fig.~\ref{fig:fork_fig}, we assess two task configurations as follows:
\begin{table}[h]
\centering
\resizebox{\linewidth}{!}{%
\begin{tabular}{l | l} 
\textit{Configuration A)} & \textit{Configuration B)} \\
$\tau_1 = \{5, 20, \text{``sen''}, \emptyset\}$,
$\tau_2 = \{7, 20, \text{``sen''}, \emptyset\}$ & 
$\tau_1 = \{5, 15, \text{``sen''}, \emptyset\}$,
$\tau_2 = \{7, 20, \text{``sen''}, \emptyset\}$ 
\\
$\tau_3 = \{5, 0, \text{``i-fus''}, [\tau_1]\}$,
$\tau_4 = \{5, 0, \text{``w-fus''}, [\tau_1, \tau_2]\}$ 
& 
$\tau_3 = \{5, 0, \text{``i-fus''}, [\tau_1]\}$,
$\tau_4 = \{5, 30, \text{``t-fus''}, [\tau_1, \tau_2]\}$
\\
$\tau_5 = \{5, 0, \text{``w-fus''}, [\tau_3, \tau_4]\}$ 
& 
$\tau_5 = \{5, 0, \text{``w-fus''}, [\tau_3, \tau_4]\}$
\\ 
\end{tabular}
}
\end{table}

In these configurations, we used a combination of T-fusion and W-fusion for $\tau_4$. However, $\tau_3$, which follows the branch node and has a single input, is modeled with I-fusion in both configurations. The results are shown in Fig.~\ref{fig:fork_results}.


\subsection{Further Evaluation with Increased Cores}\label{sec:eval-cores}
\begin{figure}[t]
    \centering
    \subfloat[Success rate]{\includegraphics[width=0.7\linewidth]{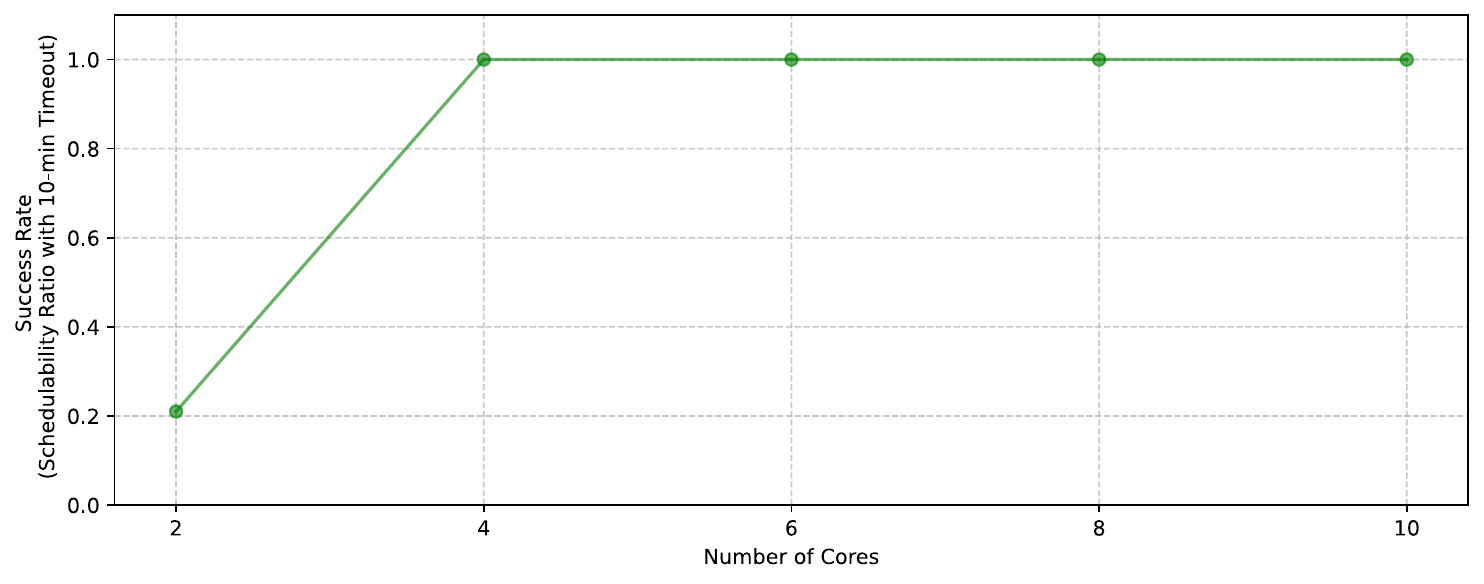}\label{fig:varying_cores_success}}
    \\ 
    \subfloat[Average runtime]{\includegraphics[width=0.7\linewidth]{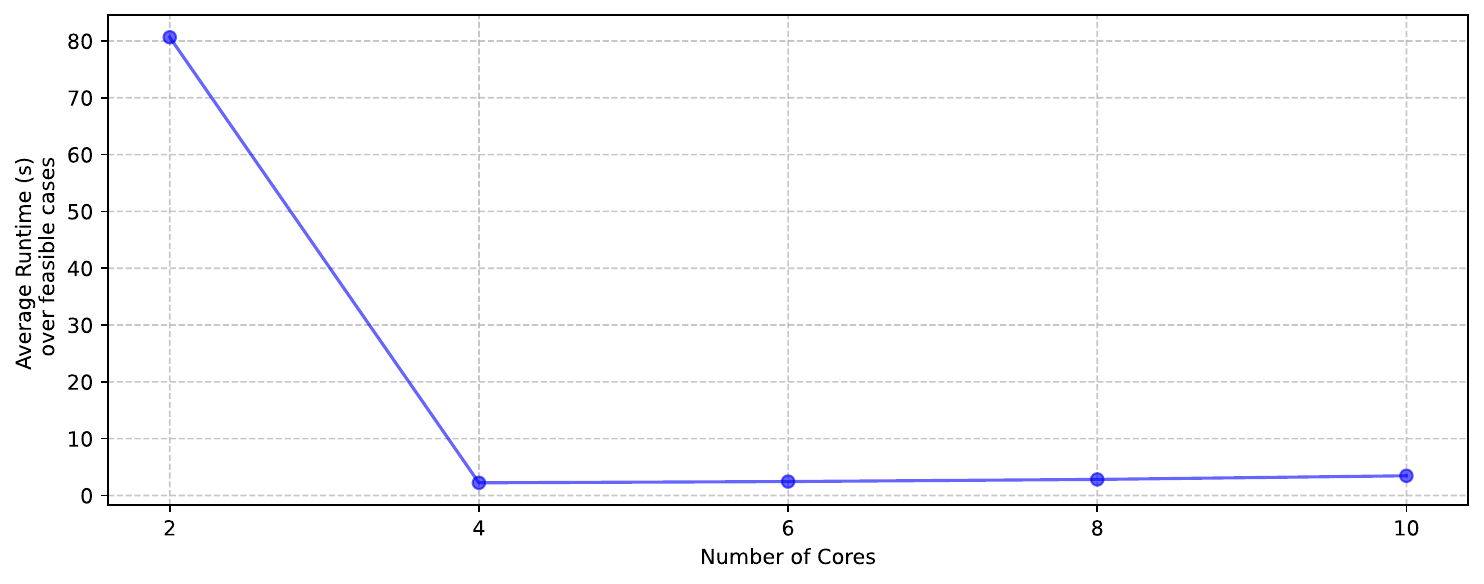}\label{fig:varying_cores_runtime}}
    \caption{The success rate \textcolor{blue}{(schedulability ratio with a 10-minute timeout)} and average runtime scaling with \#cores (\#nodes:14).}
    \label{fig:varying_cores}
\end{figure}

Although Fig.~\ref{fig:success_runtime} might initially suggest that \textcolor{blue}{the schedulability ratio (with a 10-minute timeout)} and average runtime of our framework degrade as the complexity of random DAGs increases, this trend is due to the number of cores being fixed at $\Pi=2$. In this section, we increase the number of cores and show in Fig.~\ref{fig:varying_cores} that, even for the configuration with 14 nodes, the success rate reaches 100\% when $\Pi=4$, and the average runtime becomes sufficiently low to support scaling our framework to even larger and more complex DAGs when adequate computational resources are available.

